\documentclass[aps,prb,superscriptaddress,footinbib,floatfix,showkeys,longbibliography]{revtex4-2}

\usepackage{setspace}
\makeatletter
\gdef\@ptsize{0} 
\makeatother
\usepackage{graphicx}  
\usepackage{subfigure}
\usepackage{dcolumn} 
\usepackage{bm}
\usepackage{amssymb}   
\usepackage{amsfonts}   
\usepackage{comment}
\usepackage[colorlinks,linkcolor=blue,anchorcolor=blue,citecolor=blue,urlcolor=blue]{hyperref}
\usepackage{xcolor} 
\usepackage{soul}
\usepackage{tabularx,booktabs,ragged2e,multirow}
\newcolumntype{C}{>{\centering\arraybackslash}X}
\usepackage{multirow}
\usepackage{amsmath}
\usepackage{mathrsfs}
\usepackage{mathcomp}
\usepackage{textcomp}
\usepackage{dsfont}
\usepackage{esint}
\usepackage{braket}
\usepackage{textgreek}
\usepackage{lipsum}
\usepackage{marvosym} 
\usepackage{centernot}
\usepackage{cancel}
\usepackage{dashrule}
\usepackage{tikz}
\usepackage[compat=1.1.0]{tikz-feynman}

\usepackage{stackengine}
\stackMath

\begin{document}

\title{Exciton-magnon splitting in van der Waals antiferromagnet MnPS$_3$ unveiled by second-harmonic generation}

\author{Ziqian Wang}
\email{ziqian.wang@riken.jp}
\affiliation{RIKEN Center for Emergent Matter Science (CEMS), Wako, Saitama 351-0198, Japan}

\author{Xiao-Xiao Zhang}
\affiliation{RIKEN Center for Emergent Matter Science (CEMS), Wako, Saitama 351-0198, Japan}

\author{Yuki Shiomi}
\affiliation{Department of Basic Science, University of Tokyo, Tokyo 153-8902, Japan}

\author{Taka-hisa Arima}
\affiliation{RIKEN Center for Emergent Matter Science (CEMS), Wako, Saitama 351-0198, Japan}
\affiliation{Department of Advanced Materials Science, University of Tokyo, Kashiwa, Chiba 277-8561, Japan}

\author{Naoto Nagaosa}
\affiliation{RIKEN Center for Emergent Matter Science (CEMS), Wako, Saitama 351-0198, Japan}

\author{Yoshinori Tokura}
\affiliation{RIKEN Center for Emergent Matter Science (CEMS), Wako, Saitama 351-0198, Japan}
\affiliation{Department of Applied Physics, University of Tokyo, Tokyo 113-8656, Japan}
\affiliation{Tokyo College, University of Tokyo, Tokyo 113-8656, Japan}

\author{Naoki Ogawa}
\affiliation{RIKEN Center for Emergent Matter Science (CEMS), Wako, Saitama 351-0198, Japan}


\newcommand{\ba}{{\bm a}}
\newcommand{\bb}{{\bm b}}
\newcommand{\bc}{{\bm c}}
\newcommand{\bd}{{\bm d}}
\newcommand{\bk}{{\bm k}}
\newcommand{\bmm}{{\bm m}}
\newcommand{\bn}{{\bm n}}
\newcommand{\br}{{\bm r}}
\newcommand{\bq}{{\bm q}}
\newcommand{\bp}{{\bm p}}
\newcommand{\bu}{{\bm u}}
\newcommand{\bv}{{\bm v}}
\newcommand{\bx}{{\bm x}}
\newcommand{\by}{{\bm y}}
\newcommand{\bz}{{\bm z}}
\newcommand{\bA}{{\bm A}}
\newcommand{\bB}{{\bm B}}
\newcommand{\bD}{{\bm D}}
\newcommand{\bE}{{\bm E}}
\newcommand{\bG}{{\bm G}}
\newcommand{\bH}{{\bm H}}
\newcommand{\bJ}{{\bm J}}
\newcommand{\bK}{{\bm K}}
\newcommand{\bL}{{\bm L}}
\newcommand{\bM}{{\bm M}}
\newcommand{\bP}{{\bm P}}
\newcommand{\bR}{{\bm R}}
\newcommand{\bS}{{\bm S}}
\newcommand{\bX}{{\bm X}}
\newcommand{\brho}{{\bm \rho}}
\newcommand{\cA}{{\mathcal A}}
\newcommand{\cB}{{\mathcal B}}
\newcommand{\cC}{{\mathcal C}}
\newcommand{\cD}{{\mathcal D}}
\newcommand{\cE}{{\mathcal E}}
\newcommand{\cG}{{\mathcal G}}
\newcommand{\cH}{{\mathcal H}}
\newcommand{\cJ}{{\mathcal J}}
\newcommand{\cK}{{\mathcal K}}
\newcommand{\cM}{{\mathcal M}}
\newcommand{\cP}{{\mathcal P}}
\newcommand{\cT}{{\mathcal T}}
\newcommand{\cV}{{\mathcal V}}
\newcommand{\fg}{{\mathfrak g}}
\newcommand{\fG}{{\mathfrak G}}
\newcommand{\fL}{{\mathfrak L}}
\newcommand{\sA}{{\mathscr A}}
\newcommand{\sG}{{\mathscr G}}
\newcommand{\bdelta}{{\bm \delta}}
\newcommand{\bgamma}{{\bm \gamma}}
\newcommand{\bGamma}{{\bm \Gamma}}
\newcommand{\beeta}{{\bm \eta}}
\newcommand{\bzero}{{\bm 0}}
\newcommand{\bOmega}{{\bm \Omega}}
\newcommand{\bsigma}{{\bm \sigma}}
\newcommand{\bUpsilon}{{\bm \Upsilon}}
\newcommand{\bcA}{{\bm {\mathcal A}}}
\newcommand{\bcB}{{\bm {\mathcal B}}}
\newcommand{\bcD}{{\bm {\mathcal D}}}
\newcommand\dd{\mathrm{d}}
\newcommand\ii{\mathrm{i}}
\newcommand\ee{\mathrm{e}}
\newcommand\zz{\mathtt{z}}
\newcommand\colonprod{\!:\!}

\makeatletter
\let\newtitle\@title
\let\newauthor\@author
\def\ExtendSymbol#1#2#3#4#5{\ext@arrow 0099{\arrowfill@#1#2#3}{#4}{#5}}
\newcommand\LongEqual[2][]{\ExtendSymbol{=}{=}{=}{#1}{#2}}
\newcommand\LongArrow[2][]{\ExtendSymbol{-}{-}{\rightarrow}{#1}{#2}}
\newcommand{\cev}[1]{\reflectbox{\ensuremath{\vec{\reflectbox{\ensuremath{#1}}}}}}
\newcommand{\red}[1]{\textcolor{red}{#1}} 
\newcommand{\blue}[1]{\textcolor{blue}{#1}} 
\newcommand{\green}[1]{\textcolor{orange}{#1}} 
\newcommand{\mytitle}[1]{\textcolor{orange}{\textit{#1}}}
\newcommand{\mycomment}[1]{} 
\newcommand{\note}[1]{ \textbf{\color{blue}#1}}
\newcommand{\warn}[1]{ \textbf{\color{red}#1}}

\makeatother

\begin{abstract}
Ultrafast and coherent generation of magnons is of great significance for high-speed antiferromagnetic spintronics. One possible route is by exciton-magnon pairwise optical excitation. To date, such exciton-magnon transitions have been studied mostly by linear optical means in a limited number of conventional three-dimensional antiferromagnets. Here we investigate this correlated transition in van der Waals antiferromagnet MnPS$_3$ by using resonant second-harmonic generation spectroscopy, a nonlinear optical probe sensitive to the symmetry of electronic and magnetic excitations. Two exciton-magnon peaks are observed, in line with the exciton-induced splitting of magnon density of states predicted by a Koster-Slater type theory, indicating significant exciton-magnon interactions. In addition, a large linear magnetoelectric effect of excitons is observed. These findings provide renewed understandings on the correlation effects in two-dimensional magnets with enhanced quasiparticle scattering cross-sections, and point to potential coherent control among charge, spin and orbital degrees of freedom in two-dimensional magnets by optical means.
\end{abstract}

\maketitle

\let\oldaddcontentsline\addcontentsline
\renewcommand{\addcontentsline}[3]{}

Recent discoveries of two-dimensional (2D) van der Waals (vdW) antiferromagnets (AFMs) \textit{TM}P$X_3$ (\textit{TM} = Fe, Co, Ni, Mn; \textit{X} = S, Se) with intralayer antiferromagnetic orders have brought antiferromagnetic spintronics to an unprecedented 2D realm with significant correlation and quantum effects \cite{Mak2019,Long2020a,Kim2018,Kang2020}. The N\'eel-type AFM MnPS$_3$ hosts many attractive physical properties \cite{Xing2019,Cheng2016,Shiomi2017,Takashima2018, Bostrom2021}. The reported linear magnetoelectric effect indicates the cross-correlation of its electronic and magnetic systems \cite{Ressouche2010}. Despite its potential as a superior 2D platform for magnon-based optospintronics, the efficient generation and control of magnons in MnPS$_3$ via ultrafast optical means remains a central issue.

Magnons have been excited through the coupling between spin and intermediary excitations such as onsite $d$-$d$ transitions in MnPS$_3$ and localized Zhang-Rice excitons in NiPS$_3$ \cite{Matthiesen2022a,Belvin2021,Afanasiev2021a}. However, such \textit{indirect} excitation has less cross-section and is typically confined to Brillouin zone center, resulting in the magnons with relatively low frequencies. On the other hand, the exciton fine structures in these vdW AFMs \cite{Kang2020,Hwangbo2021,Wang2021b,Ergecen2022,Gnatchenko2011} manifest the fundamental role of quasiparticle correlations. In particular, the exciton-magnon coupling revealed by the magnon-sideband in MnPS$_3$ \cite{Gnatchenko2011} suggests a possible route of ultrafast photogeneration of magnons carrying wavevectors throughout Brillouin zone, by \textit{direct} pairwise photoexcitation of a magnon and a counterpropagating exciton. However, up to now, a fundamental understanding of the magnon properties under such pair-excitation in vdW AFMs has been elusive.

Optical second-harmonic generation (SHG) is powerful for probing the magnetic symmetry in \textit{TM}PS$_3$ \cite{Chu2020,Ni2021,Ni2021d,Ni2022,Shan2021}. Using resonant SHG with symmetry analysis, we unveiled the substantial exciton-magnon interaction in MnPS$_3$, as demonstrated by the identification of two magnon bands split by excitonic perturbations, not observed in conventional 3D AFMs. The observed magnon splitting is in good consistency with the prediction by our theory of exciton-perturbed magnons. Furthermore, a large field-enhanced SHG at exciton resonance is revealed, attributed to an excitonic linear magnetoelectric effect. These findings provide renewed understandings on the correlation among quasiparticles in (quasi-)2D AFMs and may lead to efficient and coherent generation of multimode magnons.

\begin{figure}[t]
\includegraphics[width=5in]{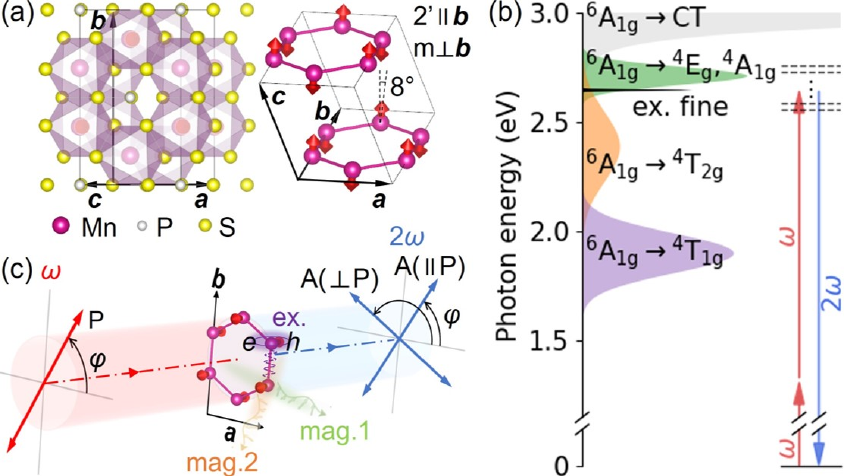}
\caption{Experimental setup for resonant SHG spectroscopy. (a) Crystal and magnetic structures of MnPS$_3$. (b) $d$-$d$ transitions and exciton fine structures shown with absorption spectrum adapted from Ref. \cite{Gnatchenko2011}. SH photon energy was set to resonate with the exciton fine structures. (c) Transmission SHG geometry with fundamental and SH light paths normal to the $ab$-plane. Optical excitation of excitons is expected to excite two types of magnons in exciton-magnon states. }\label{Fig_1}
\end{figure}

\mytitle{Exciton fine structures in SHG spectra}.--
MnPS$_3$ has a monoclinic crystal structure with space group $C2/m$ (Fig. \ref{Fig_1}(a)) \cite{Ressouche2010,Wildes2006}. In the antiferromagnetic phase ($T_{\text{N}}=78\text{ K}$), the magnetic moments of Mn$^{2+}$ ions ($S = 5/2$) align at $8^{\circ}$ to $\bc^{*}$ in the $ac$-plane ($\bc^{*}\perp \ba,\bb$), with antiferromagnetic coupling between nearest neighbors, leading to the magnetic space group $C2^{\prime}/m$ \cite{Ressouche2010,Hicks2019}. The octahedral ($O_h$) crystal field splits Mn$^{2+}$ $3d^5$ high-spin state into multiple levels, including $^{6}A_{1g}$, $^{4}T_{1g}$, $^{4}T_{2g}$, and degenerate ($^{4}E_{g}$, $^{4}A_{1g}$), among others, with observed exciton fine structures for the $^{6}A_{1g}\rightarrow {^{4}E_{g}}, {^{4}A_{1g}}$ transition (Fig. \ref{Fig_1}(b)) \cite{Gnatchenko2011}. In our experiments, the incident fundamental wavelength was adjusted such that the SH photon energy ($2 \hbar \omega$) resonates with the fine structures (Fig. \ref{Fig_1}(b)) to reveal the symmetry/anisotropy of the excited states. A transmission geometry was employed, with the incident polarization (P) and the generated SH polarization (through an analyzer, A) defined by angle $\varphi$ with respect to $a$-axis (Fig. \ref{Fig_1}(c)). 

\begin{figure}[b]
\includegraphics[width=5in]{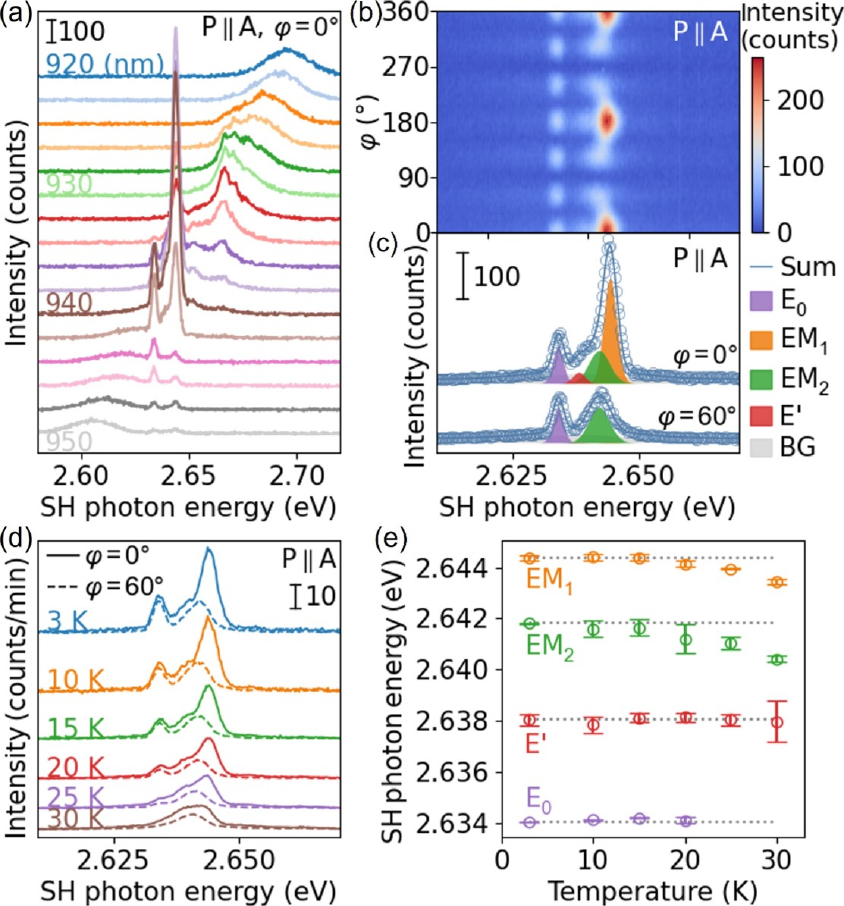}
\caption{Exciton fine structures in resonant SHG spectra. (a) Fundamental wavelength dependence of SHG spectra acquired at 2 nm intervals. 940 nm was chosen for the subsequent measurements. (b) Incident polarization (P) dependence of SHG spectra with the analyzer (A) set parallel to P. (c) SHG spectra at $\varphi=0^{\circ}$ and $60^{\circ}$ with their components indicated. Four peaks E$_0$, EM$_1$, EM$_2$, E$^\prime$ and one background BG were resolved. Spectra in (a-c) were acquired at 3 K. (d) Temperature dependence of SHG spectra at $\varphi=0^{\circ}$ and $60^{\circ}$. (e) Temperature dependence of E$_0$, EM$_1$, EM$_2$, E$^\prime$ energy extracted from (d).}\label{Fig_2}
\end{figure}

Figure \ref{Fig_2}(a) shows the SHG spectra acquired by varying the incident wavelength from 920 to 950 nm (1.348 to 1.305 eV) with P$\parallel$A, $\varphi=0^{\circ}$ at 3 K. No signal from these exciton fine structures was observed when they were off resonance, suggesting that the sharp peaks are purely from coherent SHG processes rather than incoherent ones such as two-photon luminescence. All the subsequent spectra were acquired at 940 nm central wavelength as it finds strong resonances for the exciton and its sideband simultaneously. Polarization dependence of the SHG spectra for P$\parallel$A is shown in Fig. \ref{Fig_2}(b). A single-peaked exciton-resonant SHG with 3-fold symmetry is observed at 2.634 eV (denoted as E$_0$). In stark contrast, the sideband at 2.638-2.648 eV contains at least two subpeaks with distinct $\varphi$-dependences. As detailed in Fig. \ref{Fig_2}(c), spectra at $\varphi=0^{\circ}$ and $60^{\circ}$ reveal two major peaks in the sideband region, named EM$_1$ and EM$_2$, centered at $\sim$2.644 and $\sim$2.642 eV, respectively, a minor peak E$^\prime$ at $\sim$2.638 eV and a weak broad background BG.

Temperature dependence of the fine structures aids in revealing their nature. Figure \ref{Fig_2}(d) shows the SHG spectra for P$\parallel$A, $\varphi=0^{\circ}$ and $60^{\circ}$ acquired at temperatures from 3 K to 30 K. Peak E$_0$ is discernable below 20 K, while the sidebands survive up to 30 K. The peak positions for E$_0$, EM$_1$, EM$_2$ and E$^\prime$ are plotted against temperature in Fig. \ref{Fig_2}(e). In MnPS$_3$, the magnon energy is known to decrease with increasing temperature due to the weakened exchange interaction \cite{Nagai1969}, while the phonon energy remains mostly unchanged (less than 0.12 meV shift below 30 K \cite{Sun2019}). Therefore, the obvious redshifts of EM$_1$ and EM$_2$ with increasing temperature reveal that they both have a magnon-sideband nature rather than a phonon-sideband. Besides, the nearly constant energy and weak intensity of E$^\prime$ indicate that it is most likely a phonon-sideband.

\mytitle{Exciton-magnon transitions and exciton-magnon interactions}.--
A bipartite AFM \textit{without} intersublattice interactions holds sublattice exciton $\ket{\text{E}^i,\bk}$ or magnon $\ket{\text{M}^i,\bk}$ eigenstates residing on a single sublattice $i$ ($=a$ or $b$) with wavevector $\bk$ \cite{Sell1967,Sell1968}. The optically observable exciton line is $\ket{\text{E}^{a(b)},\bm{0}}$, and the magnon-sideband is an assembly of exciton-magnon states $\ket{\text{E}^{a(b)},+\bk;\text{M}^{b(a)},-\bk}$ (and that with the sign of $\bk$ flipped) formed by an exciton and a magnon on different sublattices with opposite internal $\bk$. For Mn$^{2+}$ ions whose $d$-$d$ transitions involve a spin deviation (from $S = 5/2$ to $S = 3/2$, Fig. \ref{Fig_1}(b)), sublattice excitons derived from single-ion excitations are parity- and spin-forbidden and only allowed by weak magnetic dipole (MD) transitions. In comparison, a compound exciton-magnon transition with canceled spin deviations on different sublattices is simultaneously spin- and electric dipole (ED)-allowed \cite{Loudon1968,Tanabe1965}. In real systems \textit{with} intersublattice interactions, exciton or magnon wavefunctions can spread over neighboring sublattices, leading to the relaxation of parity/spin-selection rules and mixing of ED/MD characters. In MnPS$_3$, an exciton-magnon transition is both ED- and MD-allowed for internal $\bk$ throughout the Brillouin zone (Supplemental Material IB). Therefore, due to the near-flat exciton dispersion (Supplemental Material IC) and assuming weak exciton-magnon interactions, the magnon-sideband is anticipated to inherit mostly the spectral features of magnon density of states (DOS) exhibiting a single peak (Fig. S3). The presence of two subpeaks EM$_1$ and EM$_2$ (Figs. \ref{Fig_2}(b) and \ref{Fig_2}(c)) thus strongly indicates a departure from the weak exciton-magnon interaction regime in MnPS$_3$, in good agreement with the two subbands predicted by our theory with exciton-magnon interactions, as explained later. 

\begin{figure}[bt]
\includegraphics[width=5in]{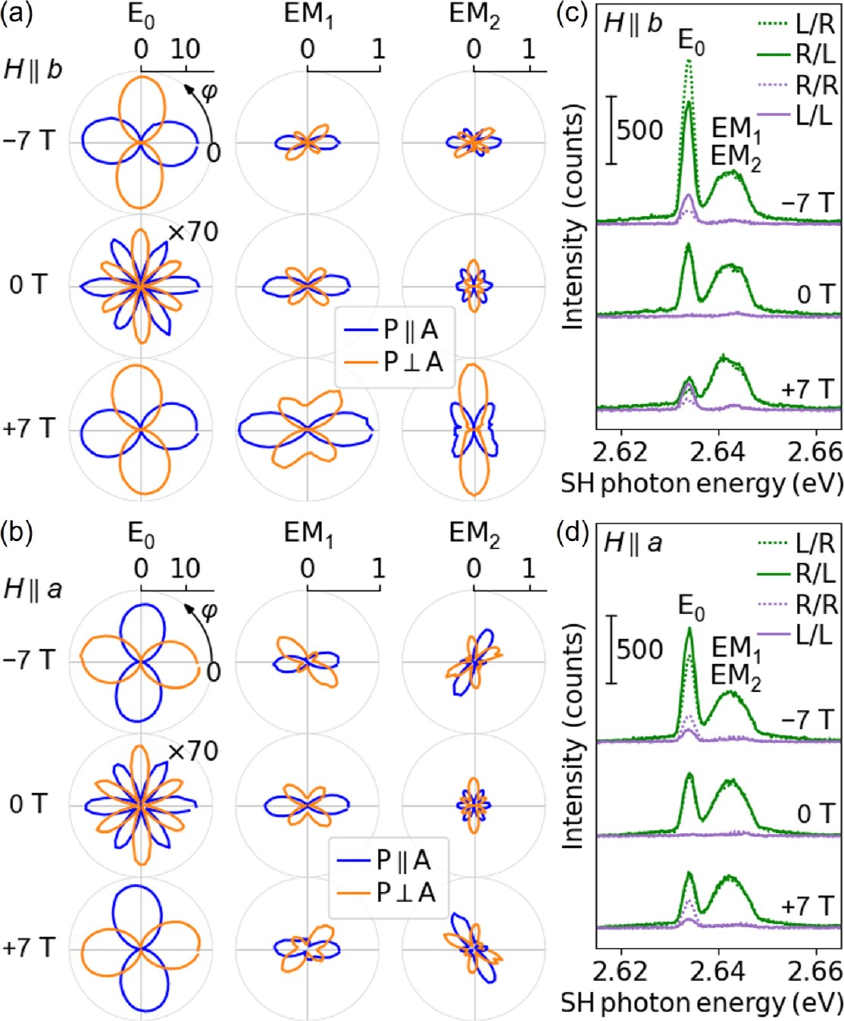}
\caption{Light polarization and magnetic field dependence of E$_0$, EM$_1$ and EM$_2$. (a,b) Linear polarization SHG-RA patterns of E$_0$, EM$_1$ and EM$_2$ under $\bH$ along $b$- and $a$-axes. Patterns in each column have a common scale bar. (c,d) Circular polarization SHG spectra under $\bH$ along $b$- and $a$-axes. L(R)/L(R): circular polarizations of fundamental/SH photons. All measurements were carried out at 3 K.}\label{Fig_3}
\end{figure}

\mytitle{Symmetry of exciton and exciton-magnon transitions}.--
To gain insights into the symmetry of the fine structures, we plot the rotational anisotropy (RA) of E$_0$, EM$_1$ and EM$_2$ under magnetic fields in Figs. \ref{Fig_3}(a) and \ref{Fig_3}(b) (see also Fig. S4 for other magnetic field conditions). Notably, E$_0$ shows a 70-fold enhancement under $\pm$7 T in both $\bH \parallel \bb$ and $\bH \parallel \ba$ orientations. At $H=0$ (Figs. \ref{Fig_3}(a) and \ref{Fig_3}(b) middle left panels), E$_0$ exhibits almost isotropic 3-fold patterns despite the broken 3-fold symmetry of magnetic space group $C2^{\prime}/m$. This near-isotropic 3-fold symmetry is consistent with the local $O_h$ crystal field of Mn$^{2+}$, thus indicating the highly localized nature of pure exciton E$_0$. In contrast, both EM$_1$ and EM$_2$ present highly anisotropic 2-fold patterns at $H=0$, which is distinct from the 3-fold local symmetry but well in line with the global $C2^{\prime}/m$ symmetry of a monolayer/bulk, suggesting the wave or extended nature of EM$_1$ and EM$_2$.

The SHG-RA patterns in Figs. \ref{Fig_3}(a) and \ref{Fig_3}(b) were further analyzed with the symmetry-adapted SH intensity formula.
\begin{widetext}
	\begin{equation}\label{eq_I_linear}
	I_{\text{SH}} \propto \left\{
	        \begin{array}{ll}
		{\left| (\chi_a \cos^2 \varphi + \chi_b \sin^2 \varphi + \chi_f \sin 2\varphi) \cos \varphi + (\chi_d \cos^2 \varphi + \chi_e \sin^2 \varphi - \chi_c \sin 2\varphi) \sin \varphi \right|}^2
		& \text{P} \parallel \text{A} \\
		{\left| (\chi_a \cos^2 \varphi + \chi_b \sin^2 \varphi + \chi_f \sin 2\varphi) \sin \varphi - (\chi_d \cos^2 \varphi + 	\chi_e \sin^2 \varphi - \chi_c \sin 2\varphi) \cos \varphi \right|}^2
		& \text{P} \perp \text{A} \\
	        \end{array}
	    \right.
	\end{equation}
\end{widetext}
Here, $\chi_a=\chi_{111}^{e}+\chi_{211}^{m}$, $\chi_b=\chi_{122}^{e}+\chi_{222}^{m}$, $\chi_c=-\chi_{212}^{e}+\chi_{112}^{m}$, $\chi_d=\chi_{211}^{e}-\chi_{111}^{m}$, $\chi_e=\chi_{222}^{e}-\chi_{122}^{m}$ and $\chi_f=\chi_{112}^{e}+\chi_{212}^{m}$, where $\chi_{ijk}^{e}$ and $\chi_{ijk}^{m}$ are the complex second-order nonlinear optical susceptibility tensors for ED and MD processes, respectively. Note that the $C2^{\prime}/m$ (for $H=0$) and $Cm$ (for $\bH \parallel \bb$) require $\chi_{d,e,f}=0$, leaving only $\chi_{a,b,c}$ active, while the $C2^{\prime}$ (for $\bH \parallel \ba$) allows all $\chi_{a,b,c,d,e,f}$. Details on the fitting method and results are given in Supplemental Material IF for reference. Linear dependences of $\chi_{a,b,c,d,e,f}$ amplitudes on $|\bH|$ were observed, suggesting the perturbative regime under these field conditions (Fig. S5). The presence of antiferromagnetic domains does not affect this analysis (Supplemental Material IF).

The symmetric behavior of E$_0$ with respect to sign reversal of $H$ in Figs. \ref{Fig_3}(a) and \ref{Fig_3}(b) (left columns) is due to the equal and linear changes in $\chi_{a,b}$ for $\bH \parallel \bb$ and in $\chi_{d,e}$ for $\bH \parallel \ba$ (Figs. S5(a) and S5(b)). This is in good agreement with the picture that the original weak E$_0$ of an MD nature at $H=0$ is easily dominated by the strong ED component emerging under the increasing $|\bH|$, leading to the 70-fold enhancement. Exemplary spectra of the significantly enhanced E$_0$ peak in magnetic fields are shown in Fig. S7. Moreover, this magnetic-field-induced ED component is polarized perpendicular to $\bH$ (Figs. \ref{Fig_3}(a) and \ref{Fig_3}(b) left columns), showing almost the same $|\bH|$ dependence for $\bH \parallel \bb$ and $\bH \parallel \ba$. Such a near-isotropic field dependence of SHG efficiency again reflects the in-plane near-isotropic nature of E$_0$ as observed in its zero-field SHG-RA patterns. 

In stark contrast, both EM$_1$ and EM$_2$ respond asymmetrically to magnetic fields (Figs. \ref{Fig_3}(a) and \ref{Fig_3}(b) middle and right columns). For $\bH \parallel \bb$, the $H$-asymmetry comes from the monotonic magnetic field dependence of $\chi_{a,b,c}$ for both EM$_1$ and EM$_2$ (Fig. S5(a)). For $\bH \parallel \ba$, $\chi_{a,b,c}$ and $\chi_f$ are not obviously affected by magnetic field, while $\chi_{d,e}$ (Fig. S5(b)) are the main cause of the $H$-asymmetric pattern distortion of EM$_1$ and EM$_2$. The coupling between magnetic field $\bH \parallel \bb$ ($\bH \parallel \ba$) and parameters $\chi_{a,b}$ ($\chi_{d,e}$) can be explained by their similar transformation properties, or the same irreducible corepresentations (ICRs), derived from group theory (Supplemental Material ID). The asymmetric magnetic field dependence suggests the interference between ED and MD processes of similar magnitudes in both EM$_1$ and EM$_2$. Intriguingly, a closer examination of the fitting parameters reveals that the relation Eq. \ref{eq_relation} is constantly satisfied in the complex numbers $\chi_{a,b,c,d,e,f}$ for EM$_1$ and EM$_2$ under all magnetic field conditions, and this is more evident in the circular polarization SHG spectra as in the following. 
\begin{equation}\label{eq_relation}
\chi_a-\chi_b \approx 2\chi_c, \quad \chi_d-\chi_e \approx 2\chi_f
\end{equation}
This relation serves as an informative clue for our symmetry analysis of the exciton-magnon states.

Figures \ref{Fig_3}(c) and \ref{Fig_3}(d) show the circular polarization SHG spectra under magnetic fields (see also Fig. S6 for other magnetic field conditions). The SH intensities in these cases are given below. 
\begin{equation}\label{eq_I_circular}
I_{\text{SH}} \propto \left\{
        \begin{array}{ll}
	{\left| (\chi_a - \chi_b - 2\chi_c) - i (\chi_d - \chi_e - 2\chi_f )\right|}^2
	& \text{L/L} \\
	{\left| (\chi_a - \chi_b - 2\chi_c) + i (\chi_d - \chi_e - 2\chi_f )\right|}^2
	& \text{R/R} \\
	{\left| (\chi_a - \chi_b + 2\chi_c) + i (\chi_d - \chi_e + 2\chi_f )\right|}^2
	& \text{L/R} \\
	{\left| (\chi_a - \chi_b + 2\chi_c) - i (\chi_d - \chi_e + 2\chi_f )\right|}^2
	& \text{R/L} \\
        \end{array}
	\right.
\end{equation}
Here, L(R)/L(R) denotes the left-handed (right-handed) circular polarization for fundamental/SH light. Again, $\chi_{d,e,f}=0$ holds for $H=0$ ($C2^{\prime}/m$) and $\bH \parallel \bb$ ($Cm$). Remarkably, for EM$_1$ and EM$_2$ (2.638-2.648 eV), L/R and R/L overlap and show almost no magnetic field dependence, and L/L and R/R signals are constantly zero. This observation confirms that Eq. \ref{eq_relation} holds individually for EM$_1$ and EM$_2$. In contrast, E$_0$ shows different peak intensities for the four polarization configurations at finite magnetic fields and exhibits distinct field dependences, thereby indicating the breakdown of Eq. \ref{eq_relation} for the pure exciton (Supplemental Material IG). It is noteworthy that the magnetic space groups $C2^{\prime}/m$, $Cm$ and $C2^{\prime}$ impose no constraints on the relation among $\chi_{a,b,c,d,e,f}$ (except for the vanishing of $\chi_{d,e,f}$ as $H$ approaches 0). Equation \ref{eq_relation} thus reflects the higher symmetry of the states in EM$_1$ and EM$_2$, which arises from the presence of equally weighted $\Gamma_1$ and $\Gamma_2$ ICR characteristics in each exciton-magnon state (Supplemental Material ID). 

\begin{figure}[b]
\includegraphics[width=5in]{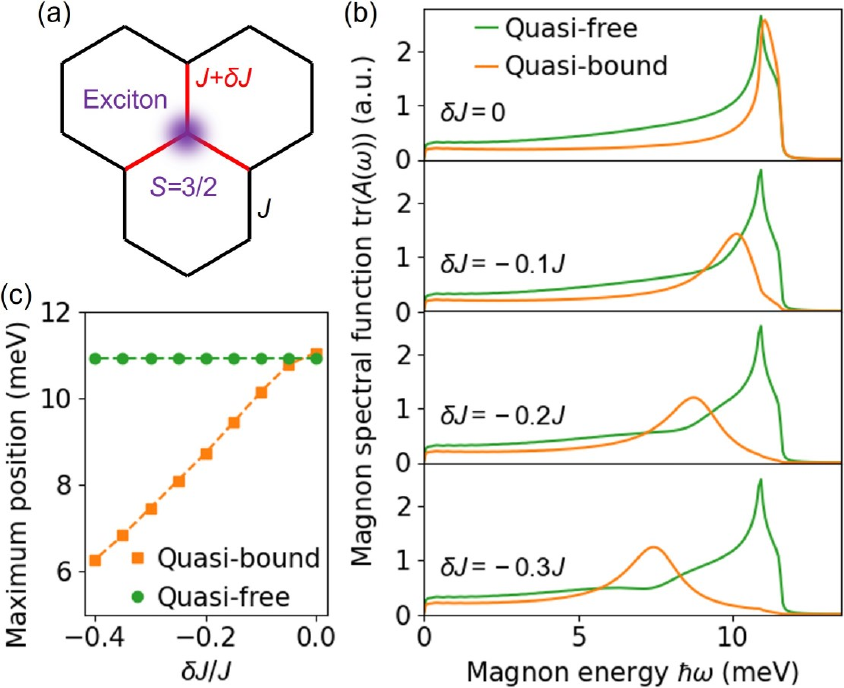}
\caption{Theoretical calculation showing exciton-induced magnon splitting in MnPS$_3$. (a) A localized exciton perturbs the spin magnitude (S) and exchange interaction ($J$). (b) Calculated magnon spectral function with varying $\delta J$. The quasi-free band shows little variation, while the quasi-bound band exhibits a significant shift and typically a broader profile. (c) $\delta J$ dependence of the energy at the maximum points of the two bands. Good agreement can be obtained between the EM$_1$/EM$_2$ in the SHG spectra (Fig. \ref{Fig_2}(c)) and the quasi-free/quasi-bound bands in the calculation at $\delta J\sim -0.25J$.}\label{Fig_4}
\end{figure}

\mytitle{Theory of exciton-perturbed magnons}.--
The overall similarities between EM$_1$ and EM$_2$ revealed by SHG suggest that they should bear similar and related physical origins, i.e., both are exciton-magnon states split from the same source. To explain the splitting, we developed a Koster-Slater type theory for magnons on a Néel AFM honeycomb lattice under the local influence due to a Frenkel exciton (See Supplemental Material II for detailed formulations). For concreteness, we conceive a physical picture of an exciton localized around an atomic site as the disturbance to the spin Hamiltonian, which changes local spin magnitude (from 5/2 to 3/2) and exchange interaction (from $J$ to $J+\delta J$) in its vicinity (Fig. \ref{Fig_4}(a)). The spin Hamiltonian $\cH=2\sum_{\braket{l,m}}{J_{lm} \bS_l\cdot\bS_m}$ only includes nearest neighbors due to their dominant contribution \cite{Wildes1998a}. 

By solving the Green’s function for the local cluster directly influenced by the exciton, two types of magnon sidebands appear. To see this, the spectral function $\text{tr}(A(\omega))$ with the sublattices summed up for the two sidebands are shown in Fig. \ref{Fig_4}(b) with varying $\delta J$ values. The spectrum nearly invariant with $\delta J$ (green) corresponds to a quasi-free magnon band, while the other with a significant $\delta J$ dependence (orange) corresponds to a quasi-bound one as the resonance occurs at an energy with finite DOS. 
The quasi-bound curve originates from the unit cell containing the localized exciton, while the quasi-free one arises from the unit cell nearby (see Supplemental Material IIB for details).
In particular, the narrower-dispersion quasi-free band has a peak position that matches the energy of EM$_1$ relative to E$_0$, while the quasi-bound band has a similar broadened feature as EM$_2$. This correlation allows for a notable correspondence between the theoretically predicted quasi-free and quasi-bound bands and the experimentally observed magnon components of EM$_1$ and EM$_2$, respectively. 

As is clear from Figs. \ref{Fig_4}(b) and \ref{Fig_4}(c), $\delta J$ is decisive to the splitting magnitude between the quasi-free and quasi-bound magnons and hence, EM$_1$ and EM$_2$. The generation of a Frenkel exciton from the Mn$^{2+}$ high-spin ground state with half-filled $d$-orbitals will create unoccupied and double-occupied orbitals, which allow ferromagnetic exchange processes with neighbors according to Goodenough-Kanamori-Anderson rules, thereby reducing the antiferromagnetic coupling in its vicinity and leading to negative $\delta J$. Comparing the theoretical spectral functions to the SHG spectra indicates that exciton causes a $\sim$25\% reduction in local exchange interaction. The substantial exciton-induced magnon splitting in MnPS$_3$, not identified in 3D AFMs, suggests a much stronger charge-spin-orbital correlation in this (quasi-)2D AFM. 

\mytitle{Discussions}.--
From a symmetry perspective, the large exciton-magnon splitting in MnPS$_3$ is associated with the shared transformation properties (ICRs) of excitons and magnons for all internal $\bk$, fully unlocking their interactions, enabled by the material’s specific site and global symmetry. This contrasts with the typical 3D AFM MnF$_2$, where exciton-magnon interactions are much suppressed due to the lack of mutual ICRs in excitons and magnons at majority high-symmetry points. A full group theoretical discussion on the magnon splitting and the related symmetry aspects of the exciton-magnon SHG is provided in Supplemental Material IB and ID.

Furthermore, as illustrated by the cartoon view in Fig. \ref{Fig_1}(c), an exciton, efficiently generated by an incident photon, simultaneously excites a magnon from either of the two magnon branches by conserving the momentum, may lead to ultrafast multimode magnon control. In addition, it is noteworthy that the magnetic-field-induced ED SHG in pure exciton E$_0$ is giant and polarized perpendicular to $\bH$, with induced SHG efficiency nearly isotropic in the plane (Figs. \ref{Fig_3}(a) and \ref{Fig_3}(b) left columns). The rotation of the exciton-originated transition dipole with $\bH$ indicates that the magnetic-field-induced electric polarization ($P_i=\alpha_{ij} H_j$, where $i,j=x,y$ and $\alpha_{xy},\alpha_{yx}$ are non-zero in-plane elements for $C2^{\prime}/m$ \cite{Ressouche2010}) explicitly occurs on the exciton, thereby leading to the \textit{excitonic} linear magnetoelectric effect. 

\nocite{Diels2006,Xu2020c,Elcoro2021,Bradley1968,Lax1961,Misetich1968,Wildes2021,Kuzemsky2016,Pantelides1978}

\begin{acknowledgments}
We thank Drs. Yuta Murakami, James Jun He, Yizhou Liu and Jannis Lehmann for valuable discussions. Z.W. was supported by JSPS KAKENHI Grant No. 21K13889. X.-X.Z was partially supported by RIKEN Special Postdoctoral Researcher Program. Y.S. was supported by JSPS KAKENHI Grant No. 22H05449. N.N. and Y.T. were supported by CREST-JST (JPMJCR1874). N.N. was supported also by JSPS KAKENHI Grant No. 18H03676. N.O. was supported by JSPS KAKENHI Grant No. 22H01185.

Z.W. and X.-X.Z. contributed equally to this work. Z.W. and N.O. designed the project and performed the SHG experiments and symmetry analysis. X.-X.Z. and N.N. performed the theoretical calculation and analysis. Y.S. grew the bulk single crystal. All authors collaborated in interpreting the data.
\end{acknowledgments}



\begin{thebibliography}{40}%
\makeatletter
\providecommand \@ifxundefined [1]{%
 \@ifx{#1\undefined}
}%
\providecommand \@ifnum [1]{%
 \ifnum #1\expandafter \@firstoftwo
 \else \expandafter \@secondoftwo
 \fi
}%
\providecommand \@ifx [1]{%
 \ifx #1\expandafter \@firstoftwo
 \else \expandafter \@secondoftwo
 \fi
}%
\providecommand \natexlab [1]{#1}%
\providecommand \enquote  [1]{``#1''}%
\providecommand \bibnamefont  [1]{#1}%
\providecommand \bibfnamefont [1]{#1}%
\providecommand \citenamefont [1]{#1}%
\providecommand \href@noop [0]{\@secondoftwo}%
\providecommand \href [0]{\begingroup \@sanitize@url \@href}%
\providecommand \@href[1]{\@@startlink{#1}\@@href}%
\providecommand \@@href[1]{\endgroup#1\@@endlink}%
\providecommand \@sanitize@url [0]{\catcode `\\12\catcode `\$12\catcode
  `\&12\catcode `\#12\catcode `\^12\catcode `\_12\catcode `\%12\relax}%
\providecommand \@@startlink[1]{}%
\providecommand \@@endlink[0]{}%
\providecommand \url  [0]{\begingroup\@sanitize@url \@url }%
\providecommand \@url [1]{\endgroup\@href {#1}{\urlprefix }}%
\providecommand \urlprefix  [0]{URL }%
\providecommand \Eprint [0]{\href }%
\providecommand \doibase [0]{https://doi.org/}%
\providecommand \selectlanguage [0]{\@gobble}%
\providecommand \bibinfo  [0]{\@secondoftwo}%
\providecommand \bibfield  [0]{\@secondoftwo}%
\providecommand \translation [1]{[#1]}%
\providecommand \BibitemOpen [0]{}%
\providecommand \bibitemStop [0]{}%
\providecommand \bibitemNoStop [0]{.\EOS\space}%
\providecommand \EOS [0]{\spacefactor3000\relax}%
\providecommand \BibitemShut  [1]{\csname bibitem#1\endcsname}%
\let\auto@bib@innerbib\@empty
\bibitem [{\citenamefont {Mak}\ \emph {et~al.}(2019)\citenamefont {Mak},
  \citenamefont {Shan},\ and\ \citenamefont {Ralph}}]{Mak2019}%
  \BibitemOpen
  \bibfield  {author} {\bibinfo {author} {\bibfnamefont {K.~F.}\ \bibnamefont
  {Mak}}, \bibinfo {author} {\bibfnamefont {J.}~\bibnamefont {Shan}},\ and\
  \bibinfo {author} {\bibfnamefont {D.~C.}\ \bibnamefont {Ralph}},\ }\bibfield
  {title} {\bibinfo {title} {{Probing and controlling magnetic states in 2D
  layered magnetic materials}},\ }\href
  {https://doi.org/10.1038/s42254-019-0110-y} {\bibfield  {journal} {\bibinfo
  {journal} {Nat. Rev. Phys.}\ }\textbf {\bibinfo {volume} {1}},\ \bibinfo
  {pages} {646} (\bibinfo {year} {2019})}\BibitemShut {NoStop}%
\bibitem [{\citenamefont {Long}\ \emph {et~al.}(2020)\citenamefont {Long},
  \citenamefont {Henck}, \citenamefont {Gibertini}, \citenamefont {Dumcenco},
  \citenamefont {Wang}, \citenamefont {Taniguchi}, \citenamefont {Watanabe},
  \citenamefont {Giannini},\ and\ \citenamefont {Morpurgo}}]{Long2020a}%
  \BibitemOpen
  \bibfield  {author} {\bibinfo {author} {\bibfnamefont {G.}~\bibnamefont
  {Long}}, \bibinfo {author} {\bibfnamefont {H.}~\bibnamefont {Henck}},
  \bibinfo {author} {\bibfnamefont {M.}~\bibnamefont {Gibertini}}, \bibinfo
  {author} {\bibfnamefont {D.}~\bibnamefont {Dumcenco}}, \bibinfo {author}
  {\bibfnamefont {Z.}~\bibnamefont {Wang}}, \bibinfo {author} {\bibfnamefont
  {T.}~\bibnamefont {Taniguchi}}, \bibinfo {author} {\bibfnamefont
  {K.}~\bibnamefont {Watanabe}}, \bibinfo {author} {\bibfnamefont
  {E.}~\bibnamefont {Giannini}},\ and\ \bibinfo {author} {\bibfnamefont
  {A.~F.}\ \bibnamefont {Morpurgo}},\ }\bibfield  {title} {\bibinfo {title}
  {{Persistence of Magnetism in Atomically Thin MnPS$_3$ Crystals}},\ }\href
  {https://doi.org/10.1021/acs.nanolett.9b05165} {\bibfield  {journal}
  {\bibinfo  {journal} {Nano Lett.}\ }\textbf {\bibinfo {volume} {20}},\
  \bibinfo {pages} {2452} (\bibinfo {year} {2020})}\BibitemShut {NoStop}%
\bibitem [{\citenamefont {Kim}\ \emph {et~al.}(2018)\citenamefont {Kim},
  \citenamefont {Kim}, \citenamefont {Sandilands}, \citenamefont {Sinn},
  \citenamefont {Lee}, \citenamefont {Son}, \citenamefont {Lee}, \citenamefont
  {Choi}, \citenamefont {Kim}, \citenamefont {Park}, \citenamefont {Jeon},
  \citenamefont {Kim}, \citenamefont {Park}, \citenamefont {Park},
  \citenamefont {Moon},\ and\ \citenamefont {Noh}}]{Kim2018}%
  \BibitemOpen
  \bibfield  {author} {\bibinfo {author} {\bibfnamefont {S.~Y.}\ \bibnamefont
  {Kim}}, \bibinfo {author} {\bibfnamefont {T.~Y.}\ \bibnamefont {Kim}},
  \bibinfo {author} {\bibfnamefont {L.~J.}\ \bibnamefont {Sandilands}},
  \bibinfo {author} {\bibfnamefont {S.}~\bibnamefont {Sinn}}, \bibinfo {author}
  {\bibfnamefont {M.~C.}\ \bibnamefont {Lee}}, \bibinfo {author} {\bibfnamefont
  {J.}~\bibnamefont {Son}}, \bibinfo {author} {\bibfnamefont {S.}~\bibnamefont
  {Lee}}, \bibinfo {author} {\bibfnamefont {K.~Y.}\ \bibnamefont {Choi}},
  \bibinfo {author} {\bibfnamefont {W.}~\bibnamefont {Kim}}, \bibinfo {author}
  {\bibfnamefont {B.~G.}\ \bibnamefont {Park}}, \bibinfo {author}
  {\bibfnamefont {C.}~\bibnamefont {Jeon}}, \bibinfo {author} {\bibfnamefont
  {H.~D.}\ \bibnamefont {Kim}}, \bibinfo {author} {\bibfnamefont {C.~H.}\
  \bibnamefont {Park}}, \bibinfo {author} {\bibfnamefont {J.~G.}\ \bibnamefont
  {Park}}, \bibinfo {author} {\bibfnamefont {S.~J.}\ \bibnamefont {Moon}},\
  and\ \bibinfo {author} {\bibfnamefont {T.~W.}\ \bibnamefont {Noh}},\
  }\bibfield  {title} {\bibinfo {title} {{Charge-Spin Correlation in van der
  Waals Antiferromagnet NiPS$_3$}},\ }\href
  {https://doi.org/10.1103/PhysRevLett.120.136402} {\bibfield  {journal}
  {\bibinfo  {journal} {Phys. Rev. Lett.}\ }\textbf {\bibinfo {volume} {120}},\
  \bibinfo {pages} {136402} (\bibinfo {year} {2018})}\BibitemShut {NoStop}%
\bibitem [{\citenamefont {Kang}\ \emph {et~al.}(2020)\citenamefont {Kang},
  \citenamefont {Kim}, \citenamefont {Kim}, \citenamefont {Kim}, \citenamefont
  {Sim}, \citenamefont {Lee}, \citenamefont {Lee}, \citenamefont {Park},
  \citenamefont {Yun}, \citenamefont {Kim}, \citenamefont {Nag}, \citenamefont
  {Walters}, \citenamefont {Garcia-Fernandez}, \citenamefont {Li},
  \citenamefont {Chapon}, \citenamefont {Zhou}, \citenamefont {Son},
  \citenamefont {Kim}, \citenamefont {Cheong},\ and\ \citenamefont
  {Park}}]{Kang2020}%
  \BibitemOpen
  \bibfield  {author} {\bibinfo {author} {\bibfnamefont {S.}~\bibnamefont
  {Kang}}, \bibinfo {author} {\bibfnamefont {K.}~\bibnamefont {Kim}}, \bibinfo
  {author} {\bibfnamefont {B.~H.}\ \bibnamefont {Kim}}, \bibinfo {author}
  {\bibfnamefont {J.}~\bibnamefont {Kim}}, \bibinfo {author} {\bibfnamefont
  {K.~I.}\ \bibnamefont {Sim}}, \bibinfo {author} {\bibfnamefont {J.-U.}\
  \bibnamefont {Lee}}, \bibinfo {author} {\bibfnamefont {S.}~\bibnamefont
  {Lee}}, \bibinfo {author} {\bibfnamefont {K.}~\bibnamefont {Park}}, \bibinfo
  {author} {\bibfnamefont {S.}~\bibnamefont {Yun}}, \bibinfo {author}
  {\bibfnamefont {T.}~\bibnamefont {Kim}}, \bibinfo {author} {\bibfnamefont
  {A.}~\bibnamefont {Nag}}, \bibinfo {author} {\bibfnamefont {A.}~\bibnamefont
  {Walters}}, \bibinfo {author} {\bibfnamefont {M.}~\bibnamefont
  {Garcia-Fernandez}}, \bibinfo {author} {\bibfnamefont {J.}~\bibnamefont
  {Li}}, \bibinfo {author} {\bibfnamefont {L.}~\bibnamefont {Chapon}}, \bibinfo
  {author} {\bibfnamefont {K.-J.}\ \bibnamefont {Zhou}}, \bibinfo {author}
  {\bibfnamefont {Y.-W.}\ \bibnamefont {Son}}, \bibinfo {author} {\bibfnamefont
  {J.~H.}\ \bibnamefont {Kim}}, \bibinfo {author} {\bibfnamefont
  {H.}~\bibnamefont {Cheong}},\ and\ \bibinfo {author} {\bibfnamefont {J.-G.}\
  \bibnamefont {Park}},\ }\bibfield  {title} {\bibinfo {title} {{Coherent
  many-body exciton in van der Waals antiferromagnet NiPS$_3$}},\ }\href
  {https://doi.org/10.1038/s41586-020-2520-5} {\bibfield  {journal} {\bibinfo
  {journal} {Nature}\ }\textbf {\bibinfo {volume} {583}},\ \bibinfo {pages}
  {785} (\bibinfo {year} {2020})}\BibitemShut {NoStop}%
\bibitem [{\citenamefont {Xing}\ \emph {et~al.}(2019)\citenamefont {Xing},
  \citenamefont {Qiu}, \citenamefont {Wang}, \citenamefont {Yao}, \citenamefont
  {Ma}, \citenamefont {Cai}, \citenamefont {Jia}, \citenamefont {Xie},\ and\
  \citenamefont {Han}}]{Xing2019}%
  \BibitemOpen
  \bibfield  {author} {\bibinfo {author} {\bibfnamefont {W.}~\bibnamefont
  {Xing}}, \bibinfo {author} {\bibfnamefont {L.}~\bibnamefont {Qiu}}, \bibinfo
  {author} {\bibfnamefont {X.}~\bibnamefont {Wang}}, \bibinfo {author}
  {\bibfnamefont {Y.}~\bibnamefont {Yao}}, \bibinfo {author} {\bibfnamefont
  {Y.}~\bibnamefont {Ma}}, \bibinfo {author} {\bibfnamefont {R.}~\bibnamefont
  {Cai}}, \bibinfo {author} {\bibfnamefont {S.}~\bibnamefont {Jia}}, \bibinfo
  {author} {\bibfnamefont {X.~C.}\ \bibnamefont {Xie}},\ and\ \bibinfo {author}
  {\bibfnamefont {W.}~\bibnamefont {Han}},\ }\bibfield  {title} {\bibinfo
  {title} {{Magnon Transport in Quasi-Two-Dimensional van der Waals
  Antiferromagnets}},\ }\href {https://doi.org/10.1103/PhysRevX.9.011026}
  {\bibfield  {journal} {\bibinfo  {journal} {Phys. Rev. X}\ }\textbf {\bibinfo
  {volume} {9}},\ \bibinfo {pages} {011026} (\bibinfo {year}
  {2019})}\BibitemShut {NoStop}%
\bibitem [{\citenamefont {Cheng}\ \emph {et~al.}(2016)\citenamefont {Cheng},
  \citenamefont {Okamoto},\ and\ \citenamefont {Xiao}}]{Cheng2016}%
  \BibitemOpen
  \bibfield  {author} {\bibinfo {author} {\bibfnamefont {R.}~\bibnamefont
  {Cheng}}, \bibinfo {author} {\bibfnamefont {S.}~\bibnamefont {Okamoto}},\
  and\ \bibinfo {author} {\bibfnamefont {D.}~\bibnamefont {Xiao}},\ }\bibfield
  {title} {\bibinfo {title} {{Spin Nernst Effect of Magnons in Collinear
  Antiferromagnets}},\ }\href {https://doi.org/10.1103/PhysRevLett.117.217202}
  {\bibfield  {journal} {\bibinfo  {journal} {Phys. Rev. Lett.}\ }\textbf
  {\bibinfo {volume} {117}},\ \bibinfo {pages} {217202} (\bibinfo {year}
  {2016})}\BibitemShut {NoStop}%
\bibitem [{\citenamefont {Shiomi}\ \emph {et~al.}(2017)\citenamefont {Shiomi},
  \citenamefont {Takashima},\ and\ \citenamefont {Saitoh}}]{Shiomi2017}%
  \BibitemOpen
  \bibfield  {author} {\bibinfo {author} {\bibfnamefont {Y.}~\bibnamefont
  {Shiomi}}, \bibinfo {author} {\bibfnamefont {R.}~\bibnamefont {Takashima}},\
  and\ \bibinfo {author} {\bibfnamefont {E.}~\bibnamefont {Saitoh}},\
  }\bibfield  {title} {\bibinfo {title} {{Experimental evidence consistent with
  a magnon Nernst effect in the antiferromagnetic insulator MnPS$_3$}},\ }\href
  {https://doi.org/10.1103/PhysRevB.96.134425} {\bibfield  {journal} {\bibinfo
  {journal} {Phys. Rev. B}\ }\textbf {\bibinfo {volume} {96}},\ \bibinfo
  {pages} {134425} (\bibinfo {year} {2017})}\BibitemShut {NoStop}%
\bibitem [{\citenamefont {Takashima}\ \emph {et~al.}(2018)\citenamefont
  {Takashima}, \citenamefont {Shiomi},\ and\ \citenamefont
  {Motome}}]{Takashima2018}%
  \BibitemOpen
  \bibfield  {author} {\bibinfo {author} {\bibfnamefont {R.}~\bibnamefont
  {Takashima}}, \bibinfo {author} {\bibfnamefont {Y.}~\bibnamefont {Shiomi}},\
  and\ \bibinfo {author} {\bibfnamefont {Y.}~\bibnamefont {Motome}},\
  }\bibfield  {title} {\bibinfo {title} {{Nonreciprocal spin Seebeck effect in
  antiferromagnets}},\ }\href {https://doi.org/10.1103/PhysRevB.98.020401}
  {\bibfield  {journal} {\bibinfo  {journal} {Phys. Rev. B}\ }\textbf {\bibinfo
  {volume} {98}},\ \bibinfo {pages} {020401(R)} (\bibinfo {year}
  {2018})}\BibitemShut {NoStop}%
\bibitem [{\citenamefont {Bostr{\"{o}}m}\ \emph {et~al.}(2021)\citenamefont
  {Bostr{\"{o}}m}, \citenamefont {Parvini}, \citenamefont {McIver},
  \citenamefont {Rubio}, \citenamefont {Kusminskiy},\ and\ \citenamefont
  {Sentef}}]{Bostrom2021}%
  \BibitemOpen
  \bibfield  {author} {\bibinfo {author} {\bibfnamefont {E.~V.}\ \bibnamefont
  {Bostr{\"{o}}m}}, \bibinfo {author} {\bibfnamefont {T.~S.}\ \bibnamefont
  {Parvini}}, \bibinfo {author} {\bibfnamefont {J.~W.}\ \bibnamefont {McIver}},
  \bibinfo {author} {\bibfnamefont {A.}~\bibnamefont {Rubio}}, \bibinfo
  {author} {\bibfnamefont {S.~V.}\ \bibnamefont {Kusminskiy}},\ and\ \bibinfo
  {author} {\bibfnamefont {M.~A.}\ \bibnamefont {Sentef}},\ }\bibfield  {title}
  {\bibinfo {title} {{All-optical generation of antiferromagnetic magnon
  currents via the magnon circular photogalvanic effect}},\ }\href
  {https://doi.org/10.1103/PhysRevB.104.L100404} {\bibfield  {journal}
  {\bibinfo  {journal} {Phys. Rev. B}\ }\textbf {\bibinfo {volume} {104}},\
  \bibinfo {pages} {L100404} (\bibinfo {year} {2021})}\BibitemShut {NoStop}%
\bibitem [{\citenamefont {Ressouche}\ \emph {et~al.}(2010)\citenamefont
  {Ressouche}, \citenamefont {Loire}, \citenamefont {Simonet}, \citenamefont
  {Ballou}, \citenamefont {Stunault},\ and\ \citenamefont
  {Wildes}}]{Ressouche2010}%
  \BibitemOpen
  \bibfield  {author} {\bibinfo {author} {\bibfnamefont {E.}~\bibnamefont
  {Ressouche}}, \bibinfo {author} {\bibfnamefont {M.}~\bibnamefont {Loire}},
  \bibinfo {author} {\bibfnamefont {V.}~\bibnamefont {Simonet}}, \bibinfo
  {author} {\bibfnamefont {R.}~\bibnamefont {Ballou}}, \bibinfo {author}
  {\bibfnamefont {A.}~\bibnamefont {Stunault}},\ and\ \bibinfo {author}
  {\bibfnamefont {A.}~\bibnamefont {Wildes}},\ }\bibfield  {title} {\bibinfo
  {title} {{Magnetoelectric MnPS$_3$ as a candidate for ferrotoroidicity}},\
  }\href {https://doi.org/10.1103/PhysRevB.82.100408} {\bibfield  {journal}
  {\bibinfo  {journal} {Phys. Rev. B}\ }\textbf {\bibinfo {volume} {82}},\
  \bibinfo {pages} {100408(R)} (\bibinfo {year} {2010})}\BibitemShut {NoStop}%
\bibitem [{\citenamefont {Matthiesen}\ \emph {et~al.}(2023)\citenamefont
  {Matthiesen}, \citenamefont {Hortensius}, \citenamefont {Ma\~{n}as Valero},
  \citenamefont {Kapon}, \citenamefont {Dumcenco}, \citenamefont {Giannini},
  \citenamefont {{\v{S}}i{\v{s}}kins}, \citenamefont {Ivanov}, \citenamefont
  {van~der Zant}, \citenamefont {Coronado}, \citenamefont {Kuzmenko},
  \citenamefont {Afanasiev},\ and\ \citenamefont {Caviglia}}]{Matthiesen2022a}%
  \BibitemOpen
  \bibfield  {author} {\bibinfo {author} {\bibfnamefont {M.}~\bibnamefont
  {Matthiesen}}, \bibinfo {author} {\bibfnamefont {J.~R.}\ \bibnamefont
  {Hortensius}}, \bibinfo {author} {\bibfnamefont {S.}~\bibnamefont {Ma\~{n}as
  Valero}}, \bibinfo {author} {\bibfnamefont {I.}~\bibnamefont {Kapon}},
  \bibinfo {author} {\bibfnamefont {D.}~\bibnamefont {Dumcenco}}, \bibinfo
  {author} {\bibfnamefont {E.}~\bibnamefont {Giannini}}, \bibinfo {author}
  {\bibfnamefont {M.}~\bibnamefont {{\v{S}}i{\v{s}}kins}}, \bibinfo {author}
  {\bibfnamefont {B.~A.}\ \bibnamefont {Ivanov}}, \bibinfo {author}
  {\bibfnamefont {H.~S.~J.}\ \bibnamefont {van~der Zant}}, \bibinfo {author}
  {\bibfnamefont {E.}~\bibnamefont {Coronado}}, \bibinfo {author}
  {\bibfnamefont {A.~B.}\ \bibnamefont {Kuzmenko}}, \bibinfo {author}
  {\bibfnamefont {D.}~\bibnamefont {Afanasiev}},\ and\ \bibinfo {author}
  {\bibfnamefont {A.~D.}\ \bibnamefont {Caviglia}},\ }\bibfield  {title}
  {\bibinfo {title} {{Controlling Magnetism with Light in a Zero Orbital
  Angular Momentum Antiferromagnet}},\ }\href
  {https://doi.org/10.1103/PhysRevLett.130.076702} {\bibfield  {journal}
  {\bibinfo  {journal} {Phys. Rev. Lett.}\ }\textbf {\bibinfo {volume} {130}},\
  \bibinfo {pages} {076702} (\bibinfo {year} {2023})}\BibitemShut {NoStop}%
\bibitem [{\citenamefont {Belvin}\ \emph {et~al.}(2021)\citenamefont {Belvin},
  \citenamefont {Baldini}, \citenamefont {Ozel}, \citenamefont {Mao},
  \citenamefont {Po}, \citenamefont {Allington}, \citenamefont {Son},
  \citenamefont {Kim}, \citenamefont {Kim}, \citenamefont {Hwang},
  \citenamefont {Kim}, \citenamefont {Park}, \citenamefont {Senthil},\ and\
  \citenamefont {Gedik}}]{Belvin2021}%
  \BibitemOpen
  \bibfield  {author} {\bibinfo {author} {\bibfnamefont {C.~A.}\ \bibnamefont
  {Belvin}}, \bibinfo {author} {\bibfnamefont {E.}~\bibnamefont {Baldini}},
  \bibinfo {author} {\bibfnamefont {I.~O.}\ \bibnamefont {Ozel}}, \bibinfo
  {author} {\bibfnamefont {D.}~\bibnamefont {Mao}}, \bibinfo {author}
  {\bibfnamefont {H.~C.}\ \bibnamefont {Po}}, \bibinfo {author} {\bibfnamefont
  {C.~J.}\ \bibnamefont {Allington}}, \bibinfo {author} {\bibfnamefont
  {S.}~\bibnamefont {Son}}, \bibinfo {author} {\bibfnamefont {B.~H.}\
  \bibnamefont {Kim}}, \bibinfo {author} {\bibfnamefont {J.}~\bibnamefont
  {Kim}}, \bibinfo {author} {\bibfnamefont {I.}~\bibnamefont {Hwang}}, \bibinfo
  {author} {\bibfnamefont {J.~H.}\ \bibnamefont {Kim}}, \bibinfo {author}
  {\bibfnamefont {J.-G.}\ \bibnamefont {Park}}, \bibinfo {author}
  {\bibfnamefont {T.}~\bibnamefont {Senthil}},\ and\ \bibinfo {author}
  {\bibfnamefont {N.}~\bibnamefont {Gedik}},\ }\bibfield  {title} {\bibinfo
  {title} {{Exciton-driven antiferromagnetic metal in a correlated van der
  Waals insulator}},\ }\href {https://doi.org/10.1038/s41467-021-25164-8}
  {\bibfield  {journal} {\bibinfo  {journal} {Nat. Commun.}\ }\textbf {\bibinfo
  {volume} {12}},\ \bibinfo {pages} {4837} (\bibinfo {year}
  {2021})}\BibitemShut {NoStop}%
\bibitem [{\citenamefont {Afanasiev}\ \emph {et~al.}(2021)\citenamefont
  {Afanasiev}, \citenamefont {Hortensius}, \citenamefont {Matthiesen},
  \citenamefont {Ma{\~{n}}as-Valero}, \citenamefont {{\v{S}}i{\v{s}}kins},
  \citenamefont {Lee}, \citenamefont {Lesne}, \citenamefont {van~der Zant},
  \citenamefont {Steeneken}, \citenamefont {Ivanov}, \citenamefont {Coronado},\
  and\ \citenamefont {Caviglia}}]{Afanasiev2021a}%
  \BibitemOpen
  \bibfield  {author} {\bibinfo {author} {\bibfnamefont {D.}~\bibnamefont
  {Afanasiev}}, \bibinfo {author} {\bibfnamefont {J.~R.}\ \bibnamefont
  {Hortensius}}, \bibinfo {author} {\bibfnamefont {M.}~\bibnamefont
  {Matthiesen}}, \bibinfo {author} {\bibfnamefont {S.}~\bibnamefont
  {Ma{\~{n}}as-Valero}}, \bibinfo {author} {\bibfnamefont {M.}~\bibnamefont
  {{\v{S}}i{\v{s}}kins}}, \bibinfo {author} {\bibfnamefont {M.}~\bibnamefont
  {Lee}}, \bibinfo {author} {\bibfnamefont {E.}~\bibnamefont {Lesne}}, \bibinfo
  {author} {\bibfnamefont {H.~S.~J.}\ \bibnamefont {van~der Zant}}, \bibinfo
  {author} {\bibfnamefont {P.~G.}\ \bibnamefont {Steeneken}}, \bibinfo {author}
  {\bibfnamefont {B.~A.}\ \bibnamefont {Ivanov}}, \bibinfo {author}
  {\bibfnamefont {E.}~\bibnamefont {Coronado}},\ and\ \bibinfo {author}
  {\bibfnamefont {A.~D.}\ \bibnamefont {Caviglia}},\ }\bibfield  {title}
  {\bibinfo {title} {{Controlling the anisotropy of a van der Waals
  antiferromagnet with light}},\ }\href
  {https://doi.org/10.1126/sciadv.abf3096} {\bibfield  {journal} {\bibinfo
  {journal} {Sci. Adv.}\ }\textbf {\bibinfo {volume} {7}},\ \bibinfo {pages}
  {eabf3096} (\bibinfo {year} {2021})}\BibitemShut {NoStop}%
\bibitem [{\citenamefont {Hwangbo}\ \emph {et~al.}(2021)\citenamefont
  {Hwangbo}, \citenamefont {Zhang}, \citenamefont {Jiang}, \citenamefont
  {Wang}, \citenamefont {Fonseca}, \citenamefont {Wang}, \citenamefont
  {Diederich}, \citenamefont {Gamelin}, \citenamefont {Xiao}, \citenamefont
  {Chu}, \citenamefont {Yao},\ and\ \citenamefont {Xu}}]{Hwangbo2021}%
  \BibitemOpen
  \bibfield  {author} {\bibinfo {author} {\bibfnamefont {K.}~\bibnamefont
  {Hwangbo}}, \bibinfo {author} {\bibfnamefont {Q.}~\bibnamefont {Zhang}},
  \bibinfo {author} {\bibfnamefont {Q.}~\bibnamefont {Jiang}}, \bibinfo
  {author} {\bibfnamefont {Y.}~\bibnamefont {Wang}}, \bibinfo {author}
  {\bibfnamefont {J.}~\bibnamefont {Fonseca}}, \bibinfo {author} {\bibfnamefont
  {C.}~\bibnamefont {Wang}}, \bibinfo {author} {\bibfnamefont {G.~M.}\
  \bibnamefont {Diederich}}, \bibinfo {author} {\bibfnamefont {D.~R.}\
  \bibnamefont {Gamelin}}, \bibinfo {author} {\bibfnamefont {D.}~\bibnamefont
  {Xiao}}, \bibinfo {author} {\bibfnamefont {J.-H.}\ \bibnamefont {Chu}},
  \bibinfo {author} {\bibfnamefont {W.}~\bibnamefont {Yao}},\ and\ \bibinfo
  {author} {\bibfnamefont {X.}~\bibnamefont {Xu}},\ }\bibfield  {title}
  {\bibinfo {title} {{Highly anisotropic excitons and multiple phonon bound
  states in a van der Waals antiferromagnetic insulator}},\ }\href
  {https://doi.org/10.1038/s41565-021-00873-9} {\bibfield  {journal} {\bibinfo
  {journal} {Nat. Nanotechnol.}\ }\textbf {\bibinfo {volume} {16}},\ \bibinfo
  {pages} {655} (\bibinfo {year} {2021})}\BibitemShut {NoStop}%
\bibitem [{\citenamefont {Wang}\ \emph {et~al.}(2021)\citenamefont {Wang},
  \citenamefont {Cao}, \citenamefont {Lu}, \citenamefont {Cohen}, \citenamefont
  {Kitadai}, \citenamefont {Li}, \citenamefont {Tan}, \citenamefont {Wilson},
  \citenamefont {Lui}, \citenamefont {Smirnov}, \citenamefont {Sharifzadeh},\
  and\ \citenamefont {Ling}}]{Wang2021b}%
  \BibitemOpen
  \bibfield  {author} {\bibinfo {author} {\bibfnamefont {X.}~\bibnamefont
  {Wang}}, \bibinfo {author} {\bibfnamefont {J.}~\bibnamefont {Cao}}, \bibinfo
  {author} {\bibfnamefont {Z.}~\bibnamefont {Lu}}, \bibinfo {author}
  {\bibfnamefont {A.}~\bibnamefont {Cohen}}, \bibinfo {author} {\bibfnamefont
  {H.}~\bibnamefont {Kitadai}}, \bibinfo {author} {\bibfnamefont
  {T.}~\bibnamefont {Li}}, \bibinfo {author} {\bibfnamefont {Q.}~\bibnamefont
  {Tan}}, \bibinfo {author} {\bibfnamefont {M.}~\bibnamefont {Wilson}},
  \bibinfo {author} {\bibfnamefont {C.~H.}\ \bibnamefont {Lui}}, \bibinfo
  {author} {\bibfnamefont {D.}~\bibnamefont {Smirnov}}, \bibinfo {author}
  {\bibfnamefont {S.}~\bibnamefont {Sharifzadeh}},\ and\ \bibinfo {author}
  {\bibfnamefont {X.}~\bibnamefont {Ling}},\ }\bibfield  {title} {\bibinfo
  {title} {{Spin-induced linear polarization of photoluminescence in
  antiferromagnetic van der Waals crystals}},\ }\href
  {https://doi.org/10.1038/s41563-021-00968-7} {\bibfield  {journal} {\bibinfo
  {journal} {Nat. Mater.}\ }\textbf {\bibinfo {volume} {20}},\ \bibinfo {pages}
  {964–} (\bibinfo {year} {2021})}\BibitemShut {NoStop}%
\bibitem [{\citenamefont {Erge{\c{c}}en}\ \emph {et~al.}(2022)\citenamefont
  {Erge{\c{c}}en}, \citenamefont {Ilyas}, \citenamefont {Mao}, \citenamefont
  {Po}, \citenamefont {Yilmaz}, \citenamefont {Kim}, \citenamefont {Park},
  \citenamefont {Senthil},\ and\ \citenamefont {Gedik}}]{Ergecen2022}%
  \BibitemOpen
  \bibfield  {author} {\bibinfo {author} {\bibfnamefont {E.}~\bibnamefont
  {Erge{\c{c}}en}}, \bibinfo {author} {\bibfnamefont {B.}~\bibnamefont
  {Ilyas}}, \bibinfo {author} {\bibfnamefont {D.}~\bibnamefont {Mao}}, \bibinfo
  {author} {\bibfnamefont {H.~C.}\ \bibnamefont {Po}}, \bibinfo {author}
  {\bibfnamefont {M.~B.}\ \bibnamefont {Yilmaz}}, \bibinfo {author}
  {\bibfnamefont {J.}~\bibnamefont {Kim}}, \bibinfo {author} {\bibfnamefont
  {J.-G.}\ \bibnamefont {Park}}, \bibinfo {author} {\bibfnamefont
  {T.}~\bibnamefont {Senthil}},\ and\ \bibinfo {author} {\bibfnamefont
  {N.}~\bibnamefont {Gedik}},\ }\bibfield  {title} {\bibinfo {title}
  {{Magnetically brightened dark electron-phonon bound states in a van der
  Waals antiferromagnet}},\ }\href {https://doi.org/10.1038/s41467-021-27741-3}
  {\bibfield  {journal} {\bibinfo  {journal} {Nat. Commun.}\ }\textbf {\bibinfo
  {volume} {13}},\ \bibinfo {pages} {98} (\bibinfo {year} {2022})}\BibitemShut
  {NoStop}%
\bibitem [{\citenamefont {Gnatchenko}\ \emph {et~al.}(2011)\citenamefont
  {Gnatchenko}, \citenamefont {Kachur}, \citenamefont {Piryatinskaya},
  \citenamefont {Vysochanskii},\ and\ \citenamefont {Gurzan}}]{Gnatchenko2011}%
  \BibitemOpen
  \bibfield  {author} {\bibinfo {author} {\bibfnamefont {S.~L.}\ \bibnamefont
  {Gnatchenko}}, \bibinfo {author} {\bibfnamefont {I.~S.}\ \bibnamefont
  {Kachur}}, \bibinfo {author} {\bibfnamefont {V.~G.}\ \bibnamefont
  {Piryatinskaya}}, \bibinfo {author} {\bibfnamefont {Y.~M.}\ \bibnamefont
  {Vysochanskii}},\ and\ \bibinfo {author} {\bibfnamefont {M.~I.}\ \bibnamefont
  {Gurzan}},\ }\bibfield  {title} {\bibinfo {title} {{Exciton-magnon structure
  of the optical absorption spectrum of antiferromagnetic MnPS$_3$}},\ }\href
  {https://doi.org/10.1063/1.3556660} {\bibfield  {journal} {\bibinfo
  {journal} {Low Temp. Phys.}\ }\textbf {\bibinfo {volume} {37}},\ \bibinfo
  {pages} {144} (\bibinfo {year} {2011})}\BibitemShut {NoStop}%
\bibitem [{\citenamefont {Chu}\ \emph {et~al.}(2020)\citenamefont {Chu},
  \citenamefont {Roh}, \citenamefont {Island}, \citenamefont {Li},
  \citenamefont {Lee}, \citenamefont {Chen}, \citenamefont {Park},
  \citenamefont {Young}, \citenamefont {Lee},\ and\ \citenamefont
  {Hsieh}}]{Chu2020}%
  \BibitemOpen
  \bibfield  {author} {\bibinfo {author} {\bibfnamefont {H.}~\bibnamefont
  {Chu}}, \bibinfo {author} {\bibfnamefont {C.~J.}\ \bibnamefont {Roh}},
  \bibinfo {author} {\bibfnamefont {J.~O.}\ \bibnamefont {Island}}, \bibinfo
  {author} {\bibfnamefont {C.}~\bibnamefont {Li}}, \bibinfo {author}
  {\bibfnamefont {S.}~\bibnamefont {Lee}}, \bibinfo {author} {\bibfnamefont
  {J.}~\bibnamefont {Chen}}, \bibinfo {author} {\bibfnamefont {J.~G.}\
  \bibnamefont {Park}}, \bibinfo {author} {\bibfnamefont {A.~F.}\ \bibnamefont
  {Young}}, \bibinfo {author} {\bibfnamefont {J.~S.}\ \bibnamefont {Lee}},\
  and\ \bibinfo {author} {\bibfnamefont {D.}~\bibnamefont {Hsieh}},\ }\bibfield
   {title} {\bibinfo {title} {{Linear Magnetoelectric Phase in Ultrathin
  MnPS$_3$ Probed by Optical Second Harmonic Generation}},\ }\href
  {https://doi.org/10.1103/PhysRevLett.124.027601} {\bibfield  {journal}
  {\bibinfo  {journal} {Phys. Rev. Lett.}\ }\textbf {\bibinfo {volume} {124}},\
  \bibinfo {pages} {027601} (\bibinfo {year} {2020})}\BibitemShut {NoStop}%
\bibitem [{\citenamefont {Ni}\ \emph {et~al.}(2021{\natexlab{a}})\citenamefont
  {Ni}, \citenamefont {Haglund}, \citenamefont {Wang}, \citenamefont {Xu},
  \citenamefont {Bernhard}, \citenamefont {Mandrus}, \citenamefont {Qian},
  \citenamefont {Mele}, \citenamefont {Kane},\ and\ \citenamefont
  {Wu}}]{Ni2021}%
  \BibitemOpen
  \bibfield  {author} {\bibinfo {author} {\bibfnamefont {Z.}~\bibnamefont
  {Ni}}, \bibinfo {author} {\bibfnamefont {A.~V.}\ \bibnamefont {Haglund}},
  \bibinfo {author} {\bibfnamefont {H.}~\bibnamefont {Wang}}, \bibinfo {author}
  {\bibfnamefont {B.}~\bibnamefont {Xu}}, \bibinfo {author} {\bibfnamefont
  {C.}~\bibnamefont {Bernhard}}, \bibinfo {author} {\bibfnamefont {D.~G.}\
  \bibnamefont {Mandrus}}, \bibinfo {author} {\bibfnamefont {X.}~\bibnamefont
  {Qian}}, \bibinfo {author} {\bibfnamefont {E.~J.}\ \bibnamefont {Mele}},
  \bibinfo {author} {\bibfnamefont {C.~L.}\ \bibnamefont {Kane}},\ and\
  \bibinfo {author} {\bibfnamefont {L.}~\bibnamefont {Wu}},\ }\bibfield
  {title} {\bibinfo {title} {{Imaging the N{\'{e}}el vector switching in the
  monolayer antiferromagnet MnPSe$_3$ with strain-controlled Ising order}},\
  }\href {https://doi.org/10.1038/s41565-021-00885-5} {\bibfield  {journal}
  {\bibinfo  {journal} {Nat. Nanotechnol.}\ }\textbf {\bibinfo {volume} {16}},\
  \bibinfo {pages} {782} (\bibinfo {year} {2021}{\natexlab{a}})}\BibitemShut
  {NoStop}%
\bibitem [{\citenamefont {Ni}\ \emph {et~al.}(2021{\natexlab{b}})\citenamefont
  {Ni}, \citenamefont {Zhang}, \citenamefont {Hopper}, \citenamefont {Haglund},
  \citenamefont {Huang}, \citenamefont {Jariwala}, \citenamefont {Bassett},
  \citenamefont {Mandrus}, \citenamefont {Mele}, \citenamefont {Kane},\ and\
  \citenamefont {Wu}}]{Ni2021d}%
  \BibitemOpen
  \bibfield  {author} {\bibinfo {author} {\bibfnamefont {Z.}~\bibnamefont
  {Ni}}, \bibinfo {author} {\bibfnamefont {H.}~\bibnamefont {Zhang}}, \bibinfo
  {author} {\bibfnamefont {D.~A.}\ \bibnamefont {Hopper}}, \bibinfo {author}
  {\bibfnamefont {A.~V.}\ \bibnamefont {Haglund}}, \bibinfo {author}
  {\bibfnamefont {N.}~\bibnamefont {Huang}}, \bibinfo {author} {\bibfnamefont
  {D.}~\bibnamefont {Jariwala}}, \bibinfo {author} {\bibfnamefont {L.~C.}\
  \bibnamefont {Bassett}}, \bibinfo {author} {\bibfnamefont {D.~G.}\
  \bibnamefont {Mandrus}}, \bibinfo {author} {\bibfnamefont {E.~J.}\
  \bibnamefont {Mele}}, \bibinfo {author} {\bibfnamefont {C.~L.}\ \bibnamefont
  {Kane}},\ and\ \bibinfo {author} {\bibfnamefont {L.}~\bibnamefont {Wu}},\
  }\bibfield  {title} {\bibinfo {title} {{Direct Imaging of Antiferromagnetic
  Domains and Anomalous Layer-Dependent Mirror Symmetry Breaking in Atomically
  Thin MnPS$_3$}},\ }\href {https://doi.org/10.1103/PhysRevLett.127.187201}
  {\bibfield  {journal} {\bibinfo  {journal} {Phys. Rev. Lett.}\ }\textbf
  {\bibinfo {volume} {127}},\ \bibinfo {pages} {187201} (\bibinfo {year}
  {2021}{\natexlab{b}})}\BibitemShut {NoStop}%
\bibitem [{\citenamefont {Ni}\ \emph {et~al.}(2022)\citenamefont {Ni},
  \citenamefont {Huang}, \citenamefont {Haglund}, \citenamefont {Mandrus},\
  and\ \citenamefont {Wu}}]{Ni2022}%
  \BibitemOpen
  \bibfield  {author} {\bibinfo {author} {\bibfnamefont {Z.}~\bibnamefont
  {Ni}}, \bibinfo {author} {\bibfnamefont {N.}~\bibnamefont {Huang}}, \bibinfo
  {author} {\bibfnamefont {A.~V.}\ \bibnamefont {Haglund}}, \bibinfo {author}
  {\bibfnamefont {D.~G.}\ \bibnamefont {Mandrus}},\ and\ \bibinfo {author}
  {\bibfnamefont {L.}~\bibnamefont {Wu}},\ }\bibfield  {title} {\bibinfo
  {title} {{Observation of Giant Surface Second-Harmonic Generation Coupled to
  Nematic Orders in the van der Waals Antiferromagnet FePS$_3$}},\ }\href
  {https://doi.org/10.1021/acs.nanolett.2c00212} {\bibfield  {journal}
  {\bibinfo  {journal} {Nano Lett.}\ }\textbf {\bibinfo {volume} {22}},\
  \bibinfo {pages} {3283} (\bibinfo {year} {2022})}\BibitemShut {NoStop}%
\bibitem [{\citenamefont {Shan}\ \emph {et~al.}(2021)\citenamefont {Shan},
  \citenamefont {Ye}, \citenamefont {Chu}, \citenamefont {Lee}, \citenamefont
  {Park}, \citenamefont {Balents},\ and\ \citenamefont {Hsieh}}]{Shan2021}%
  \BibitemOpen
  \bibfield  {author} {\bibinfo {author} {\bibfnamefont {J.-Y.}\ \bibnamefont
  {Shan}}, \bibinfo {author} {\bibfnamefont {M.}~\bibnamefont {Ye}}, \bibinfo
  {author} {\bibfnamefont {H.}~\bibnamefont {Chu}}, \bibinfo {author}
  {\bibfnamefont {S.}~\bibnamefont {Lee}}, \bibinfo {author} {\bibfnamefont
  {J.-G.}\ \bibnamefont {Park}}, \bibinfo {author} {\bibfnamefont
  {L.}~\bibnamefont {Balents}},\ and\ \bibinfo {author} {\bibfnamefont
  {D.}~\bibnamefont {Hsieh}},\ }\bibfield  {title} {\bibinfo {title} {{Giant
  modulation of optical nonlinearity by Floquet engineering}},\ }\href
  {https://doi.org/10.1038/s41586-021-04051-8} {\bibfield  {journal} {\bibinfo
  {journal} {Nature}\ }\textbf {\bibinfo {volume} {600}},\ \bibinfo {pages}
  {235} (\bibinfo {year} {2021})}\BibitemShut {NoStop}%
\bibitem [{\citenamefont {Wildes}\ \emph {et~al.}(2006)\citenamefont {Wildes},
  \citenamefont {R{\o}nnow}, \citenamefont {Roessli}, \citenamefont {Harris},\
  and\ \citenamefont {Godfrey}}]{Wildes2006}%
  \BibitemOpen
  \bibfield  {author} {\bibinfo {author} {\bibfnamefont {A.~R.}\ \bibnamefont
  {Wildes}}, \bibinfo {author} {\bibfnamefont {H.~M.}\ \bibnamefont
  {R{\o}nnow}}, \bibinfo {author} {\bibfnamefont {B.}~\bibnamefont {Roessli}},
  \bibinfo {author} {\bibfnamefont {M.~J.}\ \bibnamefont {Harris}},\ and\
  \bibinfo {author} {\bibfnamefont {K.~W.}\ \bibnamefont {Godfrey}},\
  }\bibfield  {title} {\bibinfo {title} {{Static and dynamic critical
  properties of the quasi-two-dimensional antiferromagnet MnPS$_3$}},\ }\href
  {https://doi.org/10.1103/PhysRevB.74.094422} {\bibfield  {journal} {\bibinfo
  {journal} {Phys. Rev. B}\ }\textbf {\bibinfo {volume} {74}},\ \bibinfo
  {pages} {094422} (\bibinfo {year} {2006})}\BibitemShut {NoStop}%
\bibitem [{\citenamefont {Hicks}\ \emph {et~al.}(2019)\citenamefont {Hicks},
  \citenamefont {Keller},\ and\ \citenamefont {Wildes}}]{Hicks2019}%
  \BibitemOpen
  \bibfield  {author} {\bibinfo {author} {\bibfnamefont {T.}~\bibnamefont
  {Hicks}}, \bibinfo {author} {\bibfnamefont {T.}~\bibnamefont {Keller}},\ and\
  \bibinfo {author} {\bibfnamefont {A.}~\bibnamefont {Wildes}},\ }\bibfield
  {title} {\bibinfo {title} {{Magnetic dipole splitting of magnon bands in a
  two dimensional antiferromagnet}},\ }\href
  {https://doi.org/10.1016/j.jmmm.2018.10.136} {\bibfield  {journal} {\bibinfo
  {journal} {J. Magn. Magn. Mater.}\ }\textbf {\bibinfo {volume} {474}},\
  \bibinfo {pages} {512} (\bibinfo {year} {2019})}\BibitemShut {NoStop}%
\bibitem [{\citenamefont {Nagai}\ and\ \citenamefont
  {Tanaka}(1969)}]{Nagai1969}%
  \BibitemOpen
  \bibfield  {author} {\bibinfo {author} {\bibfnamefont {O.}~\bibnamefont
  {Nagai}}\ and\ \bibinfo {author} {\bibfnamefont {T.}~\bibnamefont {Tanaka}},\
  }\bibfield  {title} {\bibinfo {title} {{Temperature-Dependent Magnon-Energy
  Theory of FeF$_2$ and MnF$_2$}},\ }\href
  {https://doi.org/10.1103/PhysRev.188.821} {\bibfield  {journal} {\bibinfo
  {journal} {Phys. Rev.}\ }\textbf {\bibinfo {volume} {188}},\ \bibinfo {pages}
  {821} (\bibinfo {year} {1969})}\BibitemShut {NoStop}%
\bibitem [{\citenamefont {Sun}\ \emph {et~al.}(2019)\citenamefont {Sun},
  \citenamefont {Tan}, \citenamefont {Liu}, \citenamefont {Gao},\ and\
  \citenamefont {Zhang}}]{Sun2019}%
  \BibitemOpen
  \bibfield  {author} {\bibinfo {author} {\bibfnamefont {Y.-J.}\ \bibnamefont
  {Sun}}, \bibinfo {author} {\bibfnamefont {Q.-H.}\ \bibnamefont {Tan}},
  \bibinfo {author} {\bibfnamefont {X.-L.}\ \bibnamefont {Liu}}, \bibinfo
  {author} {\bibfnamefont {Y.-F.}\ \bibnamefont {Gao}},\ and\ \bibinfo {author}
  {\bibfnamefont {J.}~\bibnamefont {Zhang}},\ }\bibfield  {title} {\bibinfo
  {title} {{Probing the Magnetic Ordering of Antiferromagnetic MnPS$_3$ by
  Raman Spectroscopy}},\ }\href {https://doi.org/10.1021/acs.jpclett.9b00758}
  {\bibfield  {journal} {\bibinfo  {journal} {J. Phys. Chem. Lett.}\ }\textbf
  {\bibinfo {volume} {10}},\ \bibinfo {pages} {3087} (\bibinfo {year}
  {2019})}\BibitemShut {NoStop}%
\bibitem [{\citenamefont {Sell}\ \emph {et~al.}(1967)\citenamefont {Sell},
  \citenamefont {Greene},\ and\ \citenamefont {White}}]{Sell1967}%
  \BibitemOpen
  \bibfield  {author} {\bibinfo {author} {\bibfnamefont {D.~D.}\ \bibnamefont
  {Sell}}, \bibinfo {author} {\bibfnamefont {R.~L.}\ \bibnamefont {Greene}},\
  and\ \bibinfo {author} {\bibfnamefont {R.~M.}\ \bibnamefont {White}},\
  }\bibfield  {title} {\bibinfo {title} {{Optical Exciton-Magnon Absorption in
  MnF$_2$}},\ }\href {https://doi.org/10.1103/PhysRev.158.489} {\bibfield
  {journal} {\bibinfo  {journal} {Phys. Rev.}\ }\textbf {\bibinfo {volume}
  {158}},\ \bibinfo {pages} {489} (\bibinfo {year} {1967})}\BibitemShut
  {NoStop}%
\bibitem [{\citenamefont {Sell}(1968)}]{Sell1968}%
  \BibitemOpen
  \bibfield  {author} {\bibinfo {author} {\bibfnamefont {D.~D.}\ \bibnamefont
  {Sell}},\ }\bibfield  {title} {\bibinfo {title} {{Review of Magnon‐Sideband
  Experiments}},\ }\href {https://doi.org/10.1063/1.1656158} {\bibfield
  {journal} {\bibinfo  {journal} {J. Appl. Phys.}\ }\textbf {\bibinfo {volume}
  {39}},\ \bibinfo {pages} {1030} (\bibinfo {year} {1968})}\BibitemShut
  {NoStop}%
\bibitem [{\citenamefont {Loudon}(1968)}]{Loudon1968}%
  \BibitemOpen
  \bibfield  {author} {\bibinfo {author} {\bibfnamefont {R.}~\bibnamefont
  {Loudon}},\ }\bibfield  {title} {\bibinfo {title} {{Theory of infra-red and
  optical spectra of antiferromagnets}},\ }\href
  {https://doi.org/10.1080/00018736800101296} {\bibfield  {journal} {\bibinfo
  {journal} {Adv. Phys.}\ }\textbf {\bibinfo {volume} {17}},\ \bibinfo {pages}
  {243} (\bibinfo {year} {1968})}\BibitemShut {NoStop}%
\bibitem [{\citenamefont {Tanabe}\ \emph {et~al.}(1965)\citenamefont {Tanabe},
  \citenamefont {Moriya},\ and\ \citenamefont {Sugano}}]{Tanabe1965}%
  \BibitemOpen
  \bibfield  {author} {\bibinfo {author} {\bibfnamefont {Y.}~\bibnamefont
  {Tanabe}}, \bibinfo {author} {\bibfnamefont {T.}~\bibnamefont {Moriya}},\
  and\ \bibinfo {author} {\bibfnamefont {S.}~\bibnamefont {Sugano}},\
  }\bibfield  {title} {\bibinfo {title} {{Magnon-induced electric dipole
  transition moment}},\ }\href {https://doi.org/10.1103/PhysRevLett.15.1023}
  {\bibfield  {journal} {\bibinfo  {journal} {Phys. Rev. Lett.}\ }\textbf
  {\bibinfo {volume} {15}},\ \bibinfo {pages} {1023} (\bibinfo {year}
  {1965})}\BibitemShut {NoStop}%
\bibitem [{\citenamefont {Wildes}\ \emph {et~al.}(1998)\citenamefont {Wildes},
  \citenamefont {Roessli}, \citenamefont {Lebech},\ and\ \citenamefont
  {Godfrey}}]{Wildes1998a}%
  \BibitemOpen
  \bibfield  {author} {\bibinfo {author} {\bibfnamefont {A.~R.}\ \bibnamefont
  {Wildes}}, \bibinfo {author} {\bibfnamefont {B.}~\bibnamefont {Roessli}},
  \bibinfo {author} {\bibfnamefont {B.}~\bibnamefont {Lebech}},\ and\ \bibinfo
  {author} {\bibfnamefont {K.~W.}\ \bibnamefont {Godfrey}},\ }\bibfield
  {title} {\bibinfo {title} {{Spin waves and the critical behaviour of the
  magnetization in MnPS$_3$}},\ }\href
  {https://doi.org/10.1088/0953-8984/10/28/020} {\bibfield  {journal} {\bibinfo
   {journal} {J. Phys. Condens. Matter}\ }\textbf {\bibinfo {volume} {10}},\
  \bibinfo {pages} {6417} (\bibinfo {year} {1998})}\BibitemShut {NoStop}%
\bibitem [{\citenamefont {Diels}\ and\ \citenamefont
  {Rudolph}(2006)}]{Diels2006}%
  \BibitemOpen
  \bibfield  {author} {\bibinfo {author} {\bibfnamefont {J.~C.}\ \bibnamefont
  {Diels}}\ and\ \bibinfo {author} {\bibfnamefont {W.}~\bibnamefont
  {Rudolph}},\ }\href {https://doi.org/10.1016/B978-0-12-215493-5.X5000-9}
  {\emph {\bibinfo {title} {{Ultrashort Laser Pulse Phenomena}}}},\ \bibinfo
  {edition} {second edi}\ ed.\ (\bibinfo  {publisher} {Acad. Press, Elsevier},\
  \bibinfo {year} {2006})\BibitemShut {NoStop}%
\bibitem [{\citenamefont {Xu}\ \emph {et~al.}(2020)\citenamefont {Xu},
  \citenamefont {Elcoro}, \citenamefont {Song}, \citenamefont {Wieder},
  \citenamefont {Vergniory}, \citenamefont {Regnault}, \citenamefont {Chen},
  \citenamefont {Felser},\ and\ \citenamefont {Bernevig}}]{Xu2020c}%
  \BibitemOpen
  \bibfield  {author} {\bibinfo {author} {\bibfnamefont {Y.}~\bibnamefont
  {Xu}}, \bibinfo {author} {\bibfnamefont {L.}~\bibnamefont {Elcoro}}, \bibinfo
  {author} {\bibfnamefont {Z.-D.}\ \bibnamefont {Song}}, \bibinfo {author}
  {\bibfnamefont {B.~J.}\ \bibnamefont {Wieder}}, \bibinfo {author}
  {\bibfnamefont {M.~G.}\ \bibnamefont {Vergniory}}, \bibinfo {author}
  {\bibfnamefont {N.}~\bibnamefont {Regnault}}, \bibinfo {author}
  {\bibfnamefont {Y.}~\bibnamefont {Chen}}, \bibinfo {author} {\bibfnamefont
  {C.}~\bibnamefont {Felser}},\ and\ \bibinfo {author} {\bibfnamefont {B.~A.}\
  \bibnamefont {Bernevig}},\ }\bibfield  {title} {\bibinfo {title}
  {{High-throughput calculations of magnetic topological materials}},\ }\href
  {https://doi.org/10.1038/s41586-020-2837-0} {\bibfield  {journal} {\bibinfo
  {journal} {Nature}\ }\textbf {\bibinfo {volume} {586}},\ \bibinfo {pages}
  {702} (\bibinfo {year} {2020})}\BibitemShut {NoStop}%
\bibitem [{\citenamefont {Elcoro}\ \emph {et~al.}(2021)\citenamefont {Elcoro},
  \citenamefont {Wieder}, \citenamefont {Song}, \citenamefont {Xu},
  \citenamefont {Bradlyn},\ and\ \citenamefont {Bernevig}}]{Elcoro2021}%
  \BibitemOpen
  \bibfield  {author} {\bibinfo {author} {\bibfnamefont {L.}~\bibnamefont
  {Elcoro}}, \bibinfo {author} {\bibfnamefont {B.~J.}\ \bibnamefont {Wieder}},
  \bibinfo {author} {\bibfnamefont {Z.}~\bibnamefont {Song}}, \bibinfo {author}
  {\bibfnamefont {Y.}~\bibnamefont {Xu}}, \bibinfo {author} {\bibfnamefont
  {B.}~\bibnamefont {Bradlyn}},\ and\ \bibinfo {author} {\bibfnamefont {B.~A.}\
  \bibnamefont {Bernevig}},\ }\bibfield  {title} {\bibinfo {title} {{Magnetic
  topological quantum chemistry}},\ }\href
  {https://doi.org/10.1038/s41467-021-26241-8} {\bibfield  {journal} {\bibinfo
  {journal} {Nat. Commun.}\ }\textbf {\bibinfo {volume} {12}},\ \bibinfo
  {pages} {5965} (\bibinfo {year} {2021})}\BibitemShut {NoStop}%
\bibitem [{\citenamefont {Bradley}\ and\ \citenamefont
  {Davies}(1968)}]{Bradley1968}%
  \BibitemOpen
  \bibfield  {author} {\bibinfo {author} {\bibfnamefont {C.~J.}\ \bibnamefont
  {Bradley}}\ and\ \bibinfo {author} {\bibfnamefont {B.~L.}\ \bibnamefont
  {Davies}},\ }\bibfield  {title} {\bibinfo {title} {{Magnetic Groups and Their
  Corepresentations}},\ }\href {https://doi.org/10.1103/RevModPhys.40.359}
  {\bibfield  {journal} {\bibinfo  {journal} {Rev. Mod. Phys.}\ }\textbf
  {\bibinfo {volume} {40}},\ \bibinfo {pages} {359} (\bibinfo {year}
  {1968})}\BibitemShut {NoStop}%
\bibitem [{\citenamefont {Lax}\ and\ \citenamefont {Hopfield}(1961)}]{Lax1961}%
  \BibitemOpen
  \bibfield  {author} {\bibinfo {author} {\bibfnamefont {M.}~\bibnamefont
  {Lax}}\ and\ \bibinfo {author} {\bibfnamefont {J.~J.}\ \bibnamefont
  {Hopfield}},\ }\bibfield  {title} {\bibinfo {title} {{Selection rules
  connecting different points in the brillouin zone}},\ }\href
  {https://doi.org/10.1103/PhysRev.124.115} {\bibfield  {journal} {\bibinfo
  {journal} {Phys. Rev.}\ }\textbf {\bibinfo {volume} {124}},\ \bibinfo {pages}
  {115} (\bibinfo {year} {1961})}\BibitemShut {NoStop}%
\bibitem [{\citenamefont {Misetich}\ and\ \citenamefont
  {Dietz}(1968)}]{Misetich1968}%
  \BibitemOpen
  \bibfield  {author} {\bibinfo {author} {\bibfnamefont {A.}~\bibnamefont
  {Misetich}}\ and\ \bibinfo {author} {\bibfnamefont {R.~E.}\ \bibnamefont
  {Dietz}},\ }\bibfield  {title} {\bibinfo {title} {{Role of Exciton Dispersion
  and Exciton‐Magnon Interactions on the Shape of Magnon Sidebands in
  Stressed MnF$_2$}},\ }\href {https://doi.org/10.1063/1.1656203} {\bibfield
  {journal} {\bibinfo  {journal} {J. Appl. Phys.}\ }\textbf {\bibinfo {volume}
  {39}},\ \bibinfo {pages} {1145} (\bibinfo {year} {1968})}\BibitemShut
  {NoStop}%
\bibitem [{\citenamefont {Wildes}\ \emph {et~al.}(2021)\citenamefont {Wildes},
  \citenamefont {Okamoto},\ and\ \citenamefont {Xiao}}]{Wildes2021}%
  \BibitemOpen
  \bibfield  {author} {\bibinfo {author} {\bibfnamefont {A.~R.}\ \bibnamefont
  {Wildes}}, \bibinfo {author} {\bibfnamefont {S.}~\bibnamefont {Okamoto}},\
  and\ \bibinfo {author} {\bibfnamefont {D.}~\bibnamefont {Xiao}},\ }\bibfield
  {title} {\bibinfo {title} {{Search for nonreciprocal magnons in MnPS$_3$}},\
  }\href {https://doi.org/10.1103/PhysRevB.103.024424} {\bibfield  {journal}
  {\bibinfo  {journal} {Phys. Rev. B}\ }\textbf {\bibinfo {volume} {103}},\
  \bibinfo {pages} {024424} (\bibinfo {year} {2021})}\BibitemShut {NoStop}%
\bibitem [{\citenamefont {Kuzemsky}(2016)}]{Kuzemsky2016}%
  \BibitemOpen
  \bibfield  {author} {\bibinfo {author} {\bibfnamefont {A.~L.}\ \bibnamefont
  {Kuzemsky}},\ }\href {https://doi.org/10.1142/10169} {\emph {\bibinfo {title}
  {Statistical Mechanics and the Physics of Many-Particle Model Systems}}}\
  (\bibinfo  {publisher} {{WORLD} {SCIENTIFIC}},\ \bibinfo {address}
  {Singapore},\ \bibinfo {year} {2016})\BibitemShut {NoStop}%
\bibitem [{\citenamefont {Pantelides}(1978)}]{Pantelides1978}%
  \BibitemOpen
  \bibfield  {author} {\bibinfo {author} {\bibfnamefont {S.~T.}\ \bibnamefont
  {Pantelides}},\ }\bibfield  {title} {\bibinfo {title} {The electronic
  structure of impurities and other point defects in semiconductors},\ }\href
  {https://doi.org/10.1103/revmodphys.50.797} {\bibfield  {journal} {\bibinfo
  {journal} {Reviews of Modern Physics}\ }\textbf {\bibinfo {volume} {50}},\
  \bibinfo {pages} {797} (\bibinfo {year} {1978})}\BibitemShut {NoStop}%
\end{thebibliography}

%

\let\addcontentsline\oldaddcontentsline

\end{document}